\documentclass[prx,showpacs,twocolumn,floatfix, groupaddress,superscriptaddress,nofootinbib]{revtex4-2}

\usepackage{amsmath,amsthm,amssymb}
\usepackage{mathtools}
\usepackage{dsfont}
\usepackage{physics,color,comment}
\usepackage{wasysym}
\usepackage{subcaption}
\usepackage{graphicx}
\usepackage{dcolumn}
\usepackage{bm}
\usepackage[hidelinks]{hyperref}
\usepackage[capitalize]{cleveref}
\usepackage[mathlines]{lineno}

\def \todo #1{\textcolor{red}{#1}}

\def \mc #1{\mathcal{#1}}

\newcommand{\dket}[1]{\lvert #1 \rangle\!\rangle}
\newcommand{\dbra}[1]{\langle\!\langle #1 \rvert}
\newcommand{\dbraket}[2]{\langle\!\langle #1 \vert #2 \rangle\!\rangle}

\newcommand{\naye}[1]{{\color{magenta} Naye: #1}}
\newcommand{\junan}[1]{{\color{red}Junan: #1}}

\newcommand{\ppa}{\text{PPA}3}
\newcommand{\srg}{\text{SR}2}
\newcommand{\noe}{\text{NOE2}}
\newcommand{\xhbac}{\text{xHBAC1}}

\newtheorem{theorem}{Theorem}
\newtheorem{definition}{Definition}
\newtheorem{corollary}{Corollary}

\begin{document}

\title{
Thermodynamic Analysis of Algorithmic Cooling Protocols: Efficiency Metrics and Improved Designs}

\author{Junan Lin}
\affiliation{National Research Council Canada, Toronto, ON, Canada}

\author{Nayeli A. Rodr\'iguez-Briones}
\affiliation{Department of Chemistry, University of California, Berkeley, California 94720, USA}
\affiliation{Miller Institute for Basic Research in Science, 468 Donner Lab, Berkeley, CA 94720, USA}

\author{Eduardo Mart\'in-Mart\'inez}
\affiliation{Institute for Quantum Computing, University of Waterloo, Waterloo, ON, N2L 3G1, Canada}
\affiliation{Department of Applied Mathematics, University of Waterloo, Waterloo, Ontario, N2L 3G1, Canada}
\affiliation{Perimeter Institute for Theoretical Physics, Waterloo, ON, N2L 2Y5, Canada}

\author{Raymond Laflamme}
\affiliation{Institute for Quantum Computing, University of Waterloo, Waterloo, ON, N2L 3G1, Canada}
\affiliation{Department of Physics and Astronomy, University of Waterloo, Waterloo, ON, N2L 3G1, Canada}
\affiliation{Perimeter Institute for Theoretical Physics, Waterloo, ON, N2L 2Y5, Canada}

\begin{abstract}

Algorithmic cooling (AC) protocols have been predominantly studied for their cooling capabilities, with limited attention paid to their thermodynamic properties. 
This work explores a novel perspective by analyzing a broad family of AC protocols from a thermodynamic standpoint. 
First, we give an in-depth review and formal classification of standard AC protocols. 
Leveraging the transfer matrix formalism, we achieve a consistent calculation of performance metrics, encompassing both cooling limits and target state evolution. 
We obtained a unification of these diverse cooling limits into a single, coherent mathematical expression, streamlining comparative analyses. 
Then, to assess the efficiency of coherent cooling protocols, we introduce two generic metrics: the coefficient of performance $K$ and the Landauer Ratio $R_L$, and establish a direct interrelation. Applying these metrics, we thoroughly evaluate selected AC protocols, highlighting their relative strengths.
Finally, we propose improved versions of AC protocols that exhibit enhanced thermodynamic performance, achieving desired target temperatures with lower work inputs. 
This research contributes to a deeper understanding of AC protocols and provides valuable insights for designing efficient cooling strategies in various applications.
\end{abstract}

\maketitle

\tableofcontents

\section{Introduction}

Cooling is an essential thermodynamic task in quantum science and technologies, with a wide range of applications that span from the exploration of quantum effects to the practical preparation of qubits for quantum computing. Indeed, to achieve successful quantum computation, a crucial prerequisite is the ability to generate highly pure qubits serving as initial states in most quantum algorithms, and auxiliary qubits that meet the fault tolerance threshold for quantum error correction~ \cite{knill2000theory,cory1998experimental,gottesman2010introduction}.
One of the most intuitive methods to attain this goal is reducing the environmental temperature surrounding the qubits, driving them toward their ground state. For instance, superconducting qubit-based quantum computing platforms utilize this approach, placing the qubits in a dilution refrigerator operating at milli-Kelvin (mK) temperatures~\cite{kjaergaard2020superconducting,pobell2007matter,krantz2019quantum}. While such systems have achieved cooling to around 30 mK, cooling down qubits at temperatures as low as 10 mK has proven to be challenging~\cite{krinner2019engineering,gumann2023ibm}.
Achieving the desired levels of qubit purity might demand more sophisticated and resource-optimized cooling methods, contributing to the energy constraints in quantum computing systems.

A promising alternative for cooling quantum systems below the temperatures achievable with conventional methods is to utilize information processing. Quantum information has inspired innovative methods for cooling physical systems at the quantum scale through systematic entropy manipulation techniques, known as algorithmic cooling (AC) protocols~ \cite{Park2016,Boykin2002,Schulman1999a,Fernandez2004Algorithmic,Rodriguez-Briones2017,Rodriguez-Briones2017a,Alhambra2019,clivaz2019unifying,rodriguez2020novel}. These AC protocols cool target qubits by interacting with ancillary systems through external manipulations, facilitating entropy extraction into a thermal bath. 
Several AC protocols have been proposed and experimentally tested on quantum hardware, demonstrating their universal applicability~\cite{Baugh2005,Soldati2022,Park2015,Ryan2008,Elias2011a}. These techniques offer a promising solution to enhance the purity of quantum states beyond the current limitations and provide insights into their relationship to quantum thermodynamics.

%
%
%
Understanding the impact of a cooling procedure on entropy and heat increase in the environment is essential for designing protocols in heat-sensitive applications and evaluating their practical performance.
However, a comprehensive understanding of the thermodynamics of AC protocols has been largely absent in the literature. While numerous AC protocols have been proposed and analyzed, they have predominantly been examined from a quantum information perspective, focusing on cooling limits and convergence speeds. Only recently did the notion of thermodynamic cost of cooling protocols started to gain attention. In the pioneering work by Taranto et al.~\cite{Taranto2023}, the authors examined the intricate relationship between three notions of resource in cooling (work, time, and control complexity), and studied conditions under which asymptotic limits are achievable. Also recently, Soldati et al.~\cite{Soldati2022} investigated the thermodynamics of a 3-qubit AC protocol utilizing the partner-pairing algorithm (PPA)
method and assessed its performance under certain noise conditions; and Bassman Oftelie et al.~\cite{oftelie2024dynamic} investigated the single-shot entropy compression, or Dynamic Cooling. The objective of this work is to establish a framework for studying and analyzing the thermodynamics of a broad family of AC protocols.


 
This paper is organized as follows: In Sections~\ref{sec_AC_review} and~\ref{sec_HBAC_performance}, we examine various AC protocols belonging to the coherent cooling family. We review their procedures, cooling limits, and target state evolution, employing the transition matrix formalism to derive essential quantities. This approach allows us to establish a unified perspective among these protocols.
Next, we delve into the thermodynamic aspects of each protocol. In Sections~\ref{sec_HBAC_efficiency} and~\ref{sec_Landauer}, we introduce two efficiency measures based on the required work and the amount of heat released, respectively. We demonstrate how to calculate these measures for a given protocol and compare previously studied protocols using both metrics. 
Our findings offer a means to compare the efficiency among different protocols, which we illustrate with an experimentally relevant example of cooling superconducting qubits under milli-Kelvin temperature.
Lastly, in Section~\ref{sec_improved_HBAC}, we present improved protocols that offer enhanced energy efficiency compared to the original proposals.


\subsection{Notation}


In this work, we often deal with composite systems. 
The subsystem numbering is typically indicated in the subscript.
To avoid confusion, we adopt the ``subscript with square bracket'' notation to represent full-system operators with non-trivial support on specific subsystems within a composite system. 
For instance, given an operator $A$ acting on a subsystem, $A_{[i]}$ is defined as:
\begin{equation}
    A_{[i]} = \mathds{1}_{1} \otimes \dots \otimes  \mathds{1}_{i-1} \otimes A \otimes \mathds{1}_{i+1} \otimes \dots \otimes \mathds{1}_{n}.
\end{equation}
where $\mathds{1}$ represents the identity operator.
In this notation, $\rho_{1}$ refers to the single-qubit density operator of qubit $1$, while $\sigma_{x,[1]}$ represents an $n$-qubit operator resulting from a tensor product of $\sigma_{x}$ acting on qubit $1$ and $\mathds{1}$ on the other subsystems.
Note that qubit numbering starts from $1$ in this work.


The Gibbs thermal state of a system with Hamiltonian $H$ at inverse temperature $\beta \coloneqq T^{-1}$ will be denoted as $\rho^{th}(\beta, H)$. Namely,
\begin{equation}
    \rho^{th}(\beta,H): = \frac{e^{-\beta H}}{\Tr[e^{-\beta H}]}.
\end{equation}

%

 This work primarily deals with diagonal states in the energy eigenbasis. Specifically in the context of a single-qubit, a diagonal state can be characterized by one parameter, which appears in the literature with various choices:

\begin{enumerate}
    \item The ground state ($\ket{0}$) population: $p$
    \item The excited state ($\ket{1}$) population: $\delta=1-p$
 \item Polarization $\epsilon$, defined as the difference between the ground and excited state populations $\epsilon=p-\delta$
  \item The dimensionless temperature $\beta \omega$, where $\omega$ denotes the energy gap between $\ket{0}$ and $\ket{1}$, by considering the qubit as a thermal state at inverse temperature $\beta$
\end{enumerate}
These symbols will be reserved for the corresponding quantities in this work.
To facilitate easy reference, we have compiled the transformations between these quantities in~\cref{tab_conversion}.

\begin{table*}[ht]
\renewcommand*{\arraystretch}{1.9}
\centering
\begin{tabular}{|c|c|c|c|c|}
\hline
 & $p$ & $\delta$ & $\epsilon$ & $\beta \omega$ \\ \hline
$p$ & - & $p = 1-\delta$ & $\displaystyle p = \frac{1+\epsilon}{2}$ & $\displaystyle p = \frac{1}{1+e^{-\beta \omega}}$ \\ [1ex] \hline
$\delta$ & $\delta = 1-p$ & - & $\displaystyle \delta = \frac{1-\epsilon}{2}$ & $\displaystyle \delta = \frac{1}{1+e^{\beta \omega}}$ \\ [1ex] \hline
$\epsilon$ & $\epsilon = 2p-1$ & $\epsilon = 1-2\delta$ & - & $\displaystyle \epsilon = \text{tanh}\frac{\beta \omega}{2}$ \\ [0.5ex] \hline
$\beta \omega$ & $\displaystyle \beta \omega = \log \frac{p}{1-p}$ & $\displaystyle \beta \omega = \log \frac{1-\delta}{\delta}$ & $\displaystyle \beta \omega = \log \frac{1+\epsilon}{1-\epsilon}$ & - \\ [0.8ex] \hline
\end{tabular}
\caption{Conversion table for parameters describing diagonal single-qubit states.}
\label{tab_conversion}
\end{table*}

For a system of $n$ independent qubits, we define the stationary Hamiltonian as follows:
\begin{equation}
    H_{\text{tot}} = \sum_{i=1}^{n} \omega_{i}\ketbra{1}{1}_{{[i]}},
\end{equation}
where $\omega_{i}\ketbra{1}{1}_{{[i]}}$ is the local single-qubit Hamiltonian for the $i$-th qubit with an energy gap $\omega_{i}$.
Note that this is a shifted version of the $Z$-type Hamiltonian, $-\frac{\omega_{i}}{2} \sigma_{z,[i]}$ where $\sigma_{z}$ is the Pauli-Z operator. 

%

Algorithmic Cooling protocols, when applied to systems with general initial states, incorporate a preliminary step that involves diagonalizing the state in the energy eigenbasis. This step is crucial for enhancing the purification performance (as detailed in~\cite{rodriguez2020novel,Rodriguez-Briones2017a}). An important feature of these AC protocols is that all subsequent steps consistently map diagonal operators to diagonal operators.  This allows a natural vector representation for operators by simply taking the diagonal elements as a vector. We will denote the vector representation from this mapping by $\dket{\cdot}$, such that the vector representing a diagonal operator $\rho$ is $\dket{\rho}$.

Matrices representing transformations will be denoted by $G$.
In the literature, these matrices are typically referred to as population transition matrices, as they entail the changes in energy level populations during a transformation.
In this form, the Hilbert-Schmidt inner product becomes an inner vector product as $  \langle A, B \rangle = \Tr[A^{\dagger} B] = \dbraket{A}{B}$
for $A$ and $B$ both diagonal, where $\dbra{A}$ is the conjugate transpose of $\dket{A}$.
In this work we often encounter expectation values in the form $\Tr[A B]$, which is equal to $\Tr[A^{\dagger} B]$ when $A$ is Hermitian.
For this reason we will sometimes write $\Tr[A B] = \dbraket{A}{B}$ when it is clear that $A$ is Hermitian.

\section{An overview of AC schemes: Background and Protocols}\label{sec_AC_review}


Ole W. Sørensen developed the earliest concepts of algorithmic cooling (AC) while investigating the relationship between entropy and the limitations on achievable states for spins~\cite{Sorensen1989,Sorensen1990,Sorensen1991}. He was the first to demonstrate how unitary dynamics enable a decrease in entropy for a subset of qubits while simultaneously increasing the entropy of the complementary qubits. His work provided bounds for what would later be known as reversible entropy compression, one of the building blocks for the algorithmic cooling protocols.


Subsequently, Schulman and Vazirani made significant contributions by presenting explicit methods to implement such entropy redistributions in the context of quantum information~\cite{Schulman1998,Schulman1999a}. Their approach involved performing reversible operations on a chain of qubits, to create a separation of cold and warm regions within the system. Using classical data compression results, they proposed a strategy that begins with $n$ spins of low polarization $\epsilon$, and by means of reversible redistribution of entropy within the system, produces $c n \epsilon^{2}$ qubits with purity order 1, where $c$ represents a constant. Hence, obtaining some final usable states on the order of unity requires initiating with $\mathcal{O}(1/\epsilon^2)$ noisy qubits, which for this first approach, poses a considerable cost for small $\epsilon$. (See Section~\ref{sec_single_shot} for the explicit form of the optimal entropy compression for a chain of qubits.) 

\subsubsection{PPA}\label{subsec:PPAprotocol}

An improved AC protocol over the original 
Schulman and Vazirani protocol (SV-cooling) was presented in 2007, bearing the name partner-pairing algorithm (PPA)~\cite{Schulman2007}. In the PPA,  the authors incorporated the ability to interact with an external heat bath in addition to unitary compressions within the system. This interaction facilitates the expulsion of entropy outside the system, thereby substantially increasing the achievable polarization while utilizing the same number of initial spins as in SV-cooling. In this context, we refer to the spins subject to cooling as the \textit{target spins}, while the spins used as thermal machines as the \textit{computational spins}.

\begin{figure}
\centering
\begin{subfigure}{0.96\linewidth}
    \includegraphics[width=\linewidth]{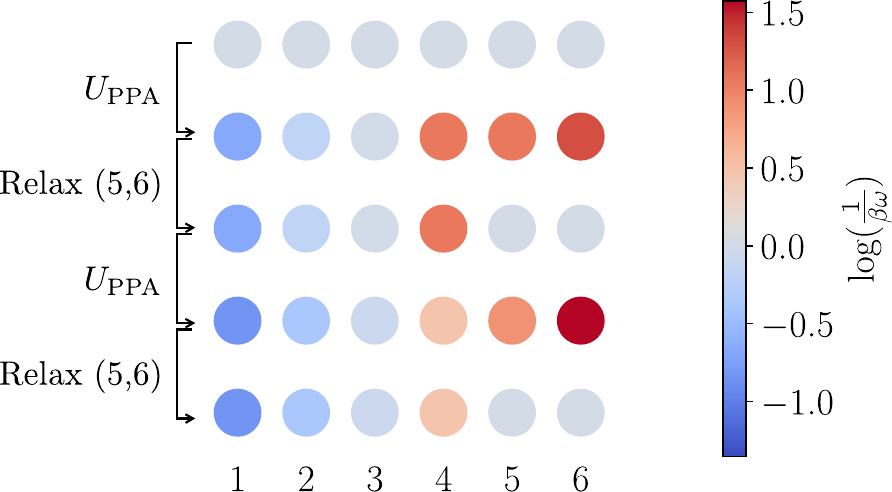}
\end{subfigure}
\caption{Visual representation of the state evolution of a string of 6 qubits under two rounds of the PPA algorithmic cooling.  The target qubit is designated as number 1, while qubits 5 and 6 serve as the reset qubits. The color coding represents the effective local temperature of each qubit, corresponding to the dimensionless temperature $(\beta \omega)^{-1}$. The process starts with all the qubits at the bath temperature $(\beta_{b} \omega)^{-1} = 1$. Then, rounds of alternating steps of entropy compression $U_\textsc{ppa}$, which concentrates the entropy towards higher index qubits, and relaxation steps which reset the last two qubits to a thermal state at bath temperature, are implemented. The compression $U_\textsc{ppa}$ implemented consists of rearranging the diagonal elements of the total density matrix of the $n$ qubits in decreasing order.
This process converges asymptotically to a cooling limit given in reference~\cite{Rodriguez-Briones2016}.}
\label{fig_PPA6q_evolution}
\end{figure}

The PPA protocol consists of repeated rounds, each involving two steps. In the first step, a global entropy compression process removes entropy from the target spins and transfers it into the computational spins, like the process in SV-cooling. This first step cools the target qubits at the expense of heating up the system's complementary part. In the second step, certain spins with higher-than-bath temperatures are refreshed through interaction with the bath, facilitating the transfer of energy from the system into the bath.
\cref{fig_PPA} illustrates one particular case of PPA cooling using 2 relaxation spins, where the relaxation step is labelled by $\Gamma_1$ and is equivalent to replacing the individual spins with a new spin at bath temperature.
\cref{fig_PPA6q_evolution} illustrates how the temperature of each qubit evolves under the PPA protocol. The system can be designed to have distinct relaxation rates for various spin types. Ideally, target spins should remain largely unchanged while computational spins rapidly equilibrate with the bath temperature. This distinction is achievable due to the diverse magnetic moments among spins, each leading to its different relaxation rate. Further details are elaborately discussed in~\cref{subsec_Gmatrix_PPA} and in references~\cite{Rodriguez-Briones2016,rodriguez2020novel}. 
Small-scale experiments with PPA have been conducted, demonstrating its ability to enhance target polarization~\cite{Baugh2005,brassard2014experimental,Soldati2022,shende2023state}.

The maximum achievable cooling using the PPA technique, which assumes that the reset step fully thermalizes the qubits to the bath temperature, is obtained when the entropy compression $U_\textsc{ppa}$ rearranges the diagonal elements of the total density matrix in decreasing order. This maximum cooling is presented and demonstrated in references~\cite{Rodriguez-Briones2016,rodriguez2020novel}. 
%

\subsubsection{NOE and SR}\label{subsec:NOESRGprotocol}

Following the discovery and effective implementation of the PPA, as well as the establishment of the protocol's cooling limits, a natural question emerged: Is it possible to go beyond the cooling limits set by the PPA?  Initially, it was believed that the PPA provided the ultimate cooling limits among the AC protocols, given its optimization considering the full thermalization of the reset spins. However, subsequent research suggested that surpassing these limits is possible by employing alternative thermalization methods.
The first protocols to demonstrate the feasibility of surpassing the PPA's cooling limits were the Nuclear Overhauser Effect Heat-Bath Algorithmic Cooling (NOE-HBAC) and the State-Reset Heat-Bath Algorithmic Cooling (SR-HBAC) protocols. Both protocols were introduced in reference~\cite{Rodriguez-Briones2017}.
These protocols, NOE-HBAC and SR-HBAC, require not just the ability to fully thermalize with a heat bath, but also the capability to selectively "thermalize" specific populations of energy levels within the composite spin system. This novel approach is influenced by the Nuclear Overhauser Effect, where cross-relaxation between nearby spins allows the polarization of one spin to increase when the other spin is saturated to the maximally mixed state.

The NOE-HBAC and SR-HBAC protocols can cool a target qubit using just a single computational qubit, which is notably distinct from the PPA that demands at least two computational qubits. Furthermore, the SR-HBAC outperforms the NOE-HBAC by achieving greater final polarization. In the circuits depicted in \cref{fig_NOE} and \cref{fig_SR2}, the SR step is represented by $\Gamma_{2}$. This indicates that a more refined relaxation pathway for individual energy levels can be essential for cooling. Further details of these protocols are elaborately discussed in~\cref{subsec_Gmatrix_NOE,subsec_Gmatrix_SR2} and in references~\cite{Rodriguez-Briones2017,rodriguez2020novel}. 




\subsubsection{xHBAC}\label{subsec:xHBACprotocol}

Recently, further generalizations of the SR-based HBAC have been proposed.
 Notably, among these new protocols, the Extended Heat-Bath Algorithmic Cooling (xHBAC) is particularly significant for its ability to asymptotically cool to a polarization of one (corresponding to zero temperature), as elaborated in Alhambra et al.~\cite{Alhambra2019}.
Within the xHBAC framework, the state-reset (SR) step was identified 
as a component of a broader category of quantum operations termed ``dephasing thermalization''. This insight revealed that enhanced cooling is achievable by performing arbitrary dephasing thermalizations and optimizing within this set of operations.
Specifically, the xHBAC protocol allows for a single qubit to be cooled exponentially fast towards a polarization of one using the optimal thermalization operation, represented by $\Lambda_{\beta}$ in \cref{fig_xHBAC_1}, and detailed in~\cref{subsec_Gmatrix_xHBAC}.
At first glance, this result may seem to challenge the second law of thermodynamics. However, it can be reconciled by considering that performing the optimal thermalization requires detailed interaction with an infinite dimensional ancillary system, which has a diverging control complexity as a resource ---being consistent with reference~\cite{Taranto2023}.

While this paper focuses on a select range of algorithmic cooling protocols, we acknowledge the existence of other cooling methods for interacting quantum systems that are not covered in this discussion, for instance the protocols presented in references~\cite{Taranto2023,taranto2020exponential,clivaz2019unifying,cotler2019quantum}. An alternative family of methods has shown considerable potential in strongly interacting systems, as detailed in reference~\cite{Rodriguez-Briones2017a}. These protocols employ local operations and classical communication to distinctively utilize correlations arising from internal interactions through quantum energy teleportation protocols~\cite{hotta2009quantum,frey2013quantum,hotta2008protocol}.

\subsection{Unified framework: Coherent control AC}\label{subsec_coherent_control}
All the AC protocols discussed above fall under the category of coherent control cooling~\cite{Taranto2023}.
The setup consists of three components: a \emph{target} system to be cooled (denoted by a subscript $t$), a thermal \emph{machine} (denoted by a subscript $m$), and an external environment acting as a heat \emph{bath} (denoted by a subscript $b$).
Collectively, the combined target and machine will simply be referred to as the \emph{system}.
Coherent control cooling protocols are composed of two basic subroutines: a unitary \textit{control} step and a \textit{thermalization} step.
The precise meanings of these two subroutines, clarifying their specific roles and functions, are given in the following definitions.

\begin{definition}
    A control subroutine is characterized by a unitary $U_{t,m}$ that acts on both the target and the machine subsystems.
\end{definition}

In the control step, a unitary is applied between the target system and the machine.
This step has two potential objectives, depending on the specific protocol in use. Its primary function is to facilitate energy transfer from the target to the machine, effectively cooling the target. Alternatively, its role could be to prepare the system's state to optimize the effect of the subsequent thermalization step. These two functions may be employed independently based on the desired outcome. 
The unitary operation can be arbitrary and does not need to preserve energy; that is, applying $U_{t,m}$ can result in a net input of work.

\begin{definition}\label{defn_thermalization}
    A thermalization subroutine is characterized by a \emph{thermal operation} (TO) map $\mc{E}$ acting on both the target and machine subsystems. Specifically, it can be expressed as
    \begin{equation}
        \mc{E}(\rho_{t,m}) = \Tr_{b} [V_{t,m,b}(\rho_{t,m} \otimes \rho^{th}(\beta_{b}, H_{b}))V^{\dagger}_{t,m,b}]
    \end{equation}
    where the joint unitary $V_{t,m,b}$ acting on the target, machine, and bath, satisfies $[V_{t,m,b}, H_{t} + H_{m} + H_{b}]=0$.
\end{definition}

In the thermalization step, part of the system is allowed to exchange energy with the heat bath, typically resulting in a net transfer of energy from the system to the bath.
When a protocol is repeated for multiple rounds, it will be assumed that the bath is sufficiently large and always starts from a thermal state, and any correlation effects potentially building up during the relaxation will be disregarded.

    The transition matrix $G_{\mc{E}}$ of any TO map $\mc{E}$ is Gibbs-stochastic~\cite{Lostaglio2018}.
Here, Gibbs-stochastic refers to a stochastic matrix that preserves a Gibbs state input, i.e. $G_{\mc{E}} \dket{\rho^{th}} = \dket{\rho^{th}}$.
This property can be seen by substituting the definition of $\rho^{th}$ into \cref{defn_thermalization} and noticing that $[V_{t,m,b}, H_{t} + H_{m} + H_{b}]=0$.

The thermalization subroutine can be realized without any external work input.
For example, resetting a qubit to the thermal state at the bath's temperature, which constitutes complete thermalization, can be achieved by simply putting a system in contact with a heat bath over a extended duration. This process does not require any work input.
Here we have adopted a broader definition for the thermalization subroutine than just complete thermalization.
These more general thermal operations are components of the cooling protocols introduced in this work.

\begin{definition}
    A Coherent Control Algorithmic Cooling (AC) protocol is defined as a methodical sequence in which each step alternates between a control subroutine, or a thermalization subroutine.
\end{definition}

\begin{figure}
\centering
\begin{subfigure}{0.64\linewidth}
    \includegraphics[width=\linewidth]{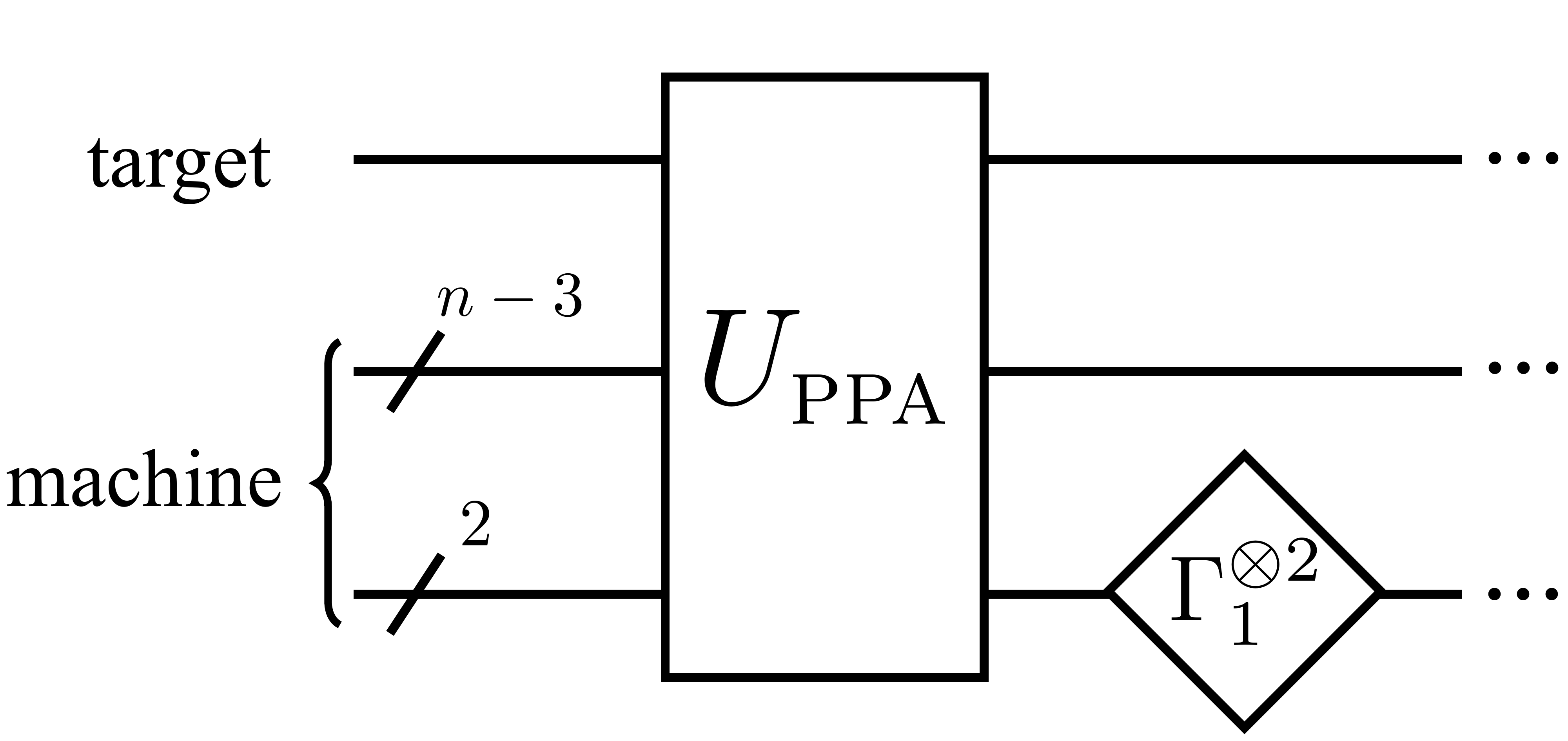}
    \caption{\textbf{Round of PPA}}
    \label{fig_PPA}
\end{subfigure}
\hfill
\begin{subfigure}{0.66\linewidth}
    \includegraphics[width=\linewidth]{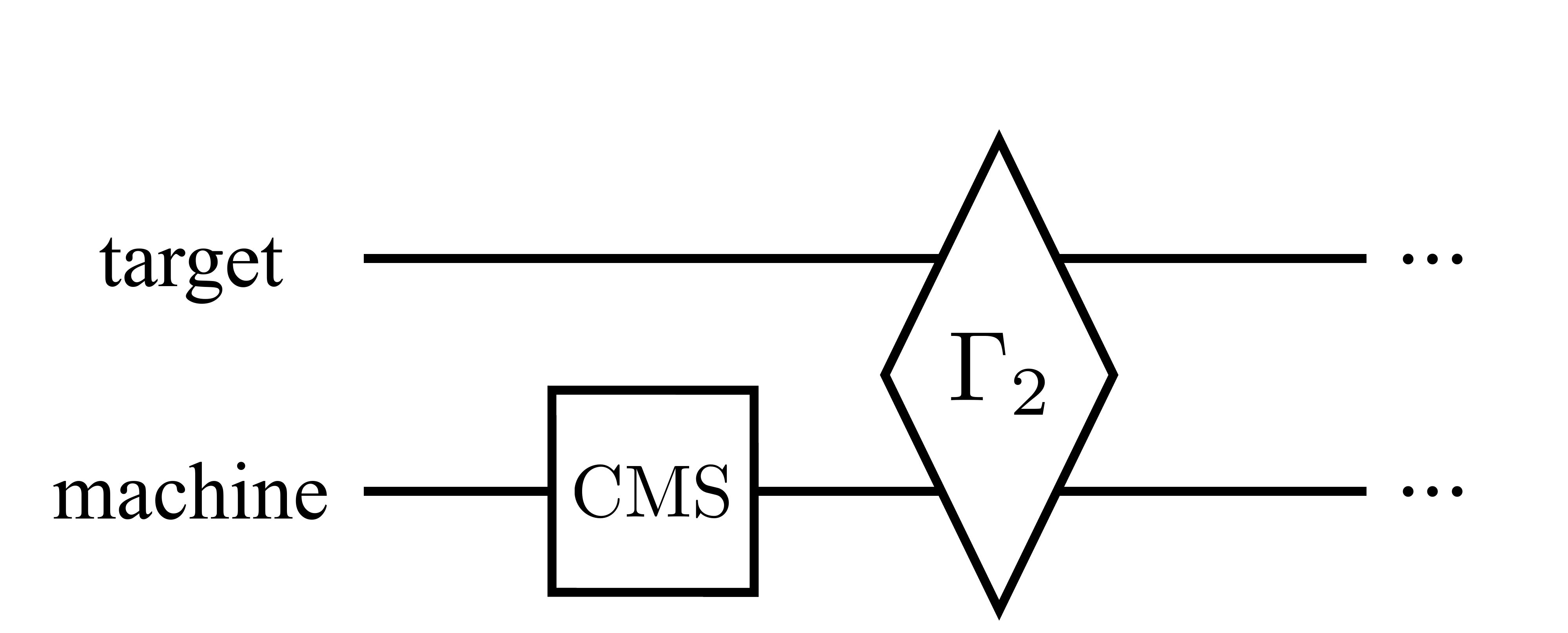}
    \caption{\textbf{Round of NOE2}}
    \label{fig_NOE}
\end{subfigure}
\hfill
\begin{subfigure}{0.66\linewidth}
    \includegraphics[width=\linewidth]{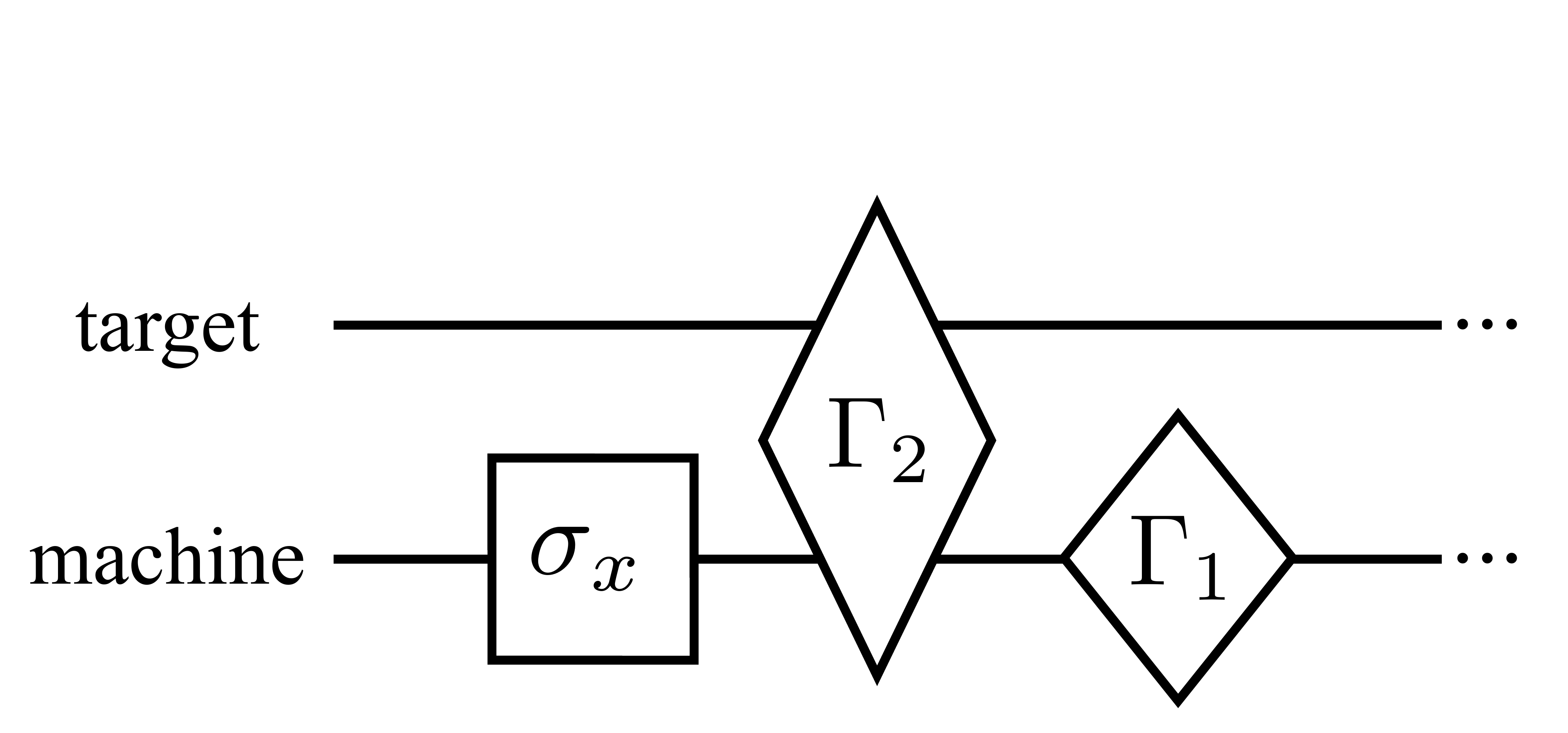}
    \caption{\textbf{Round of SR2}}
    \label{fig_SR2}
\end{subfigure}
\hfill
\begin{subfigure}{0.66\linewidth}
    \includegraphics[width=\linewidth]{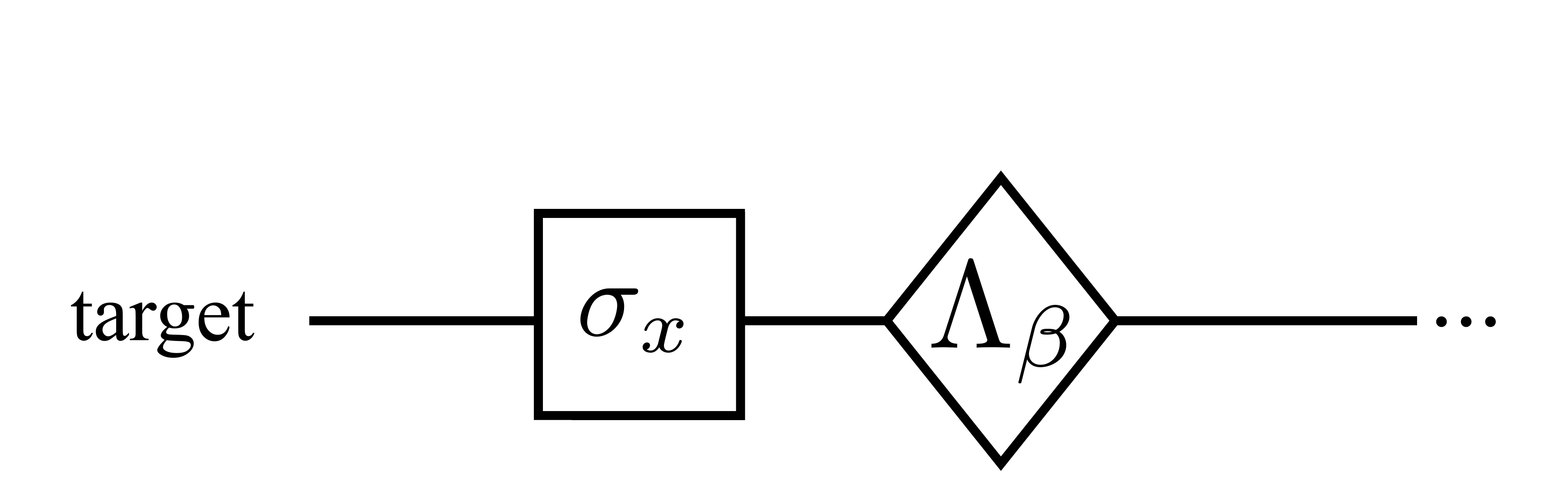}
    \caption{\textbf{Round of xHBAC1}}
    \label{fig_xHBAC_1}
\end{subfigure}
       
\caption{\textbf{Circuit diagrams illustrating the rounds of HBAC protocols.} Rectangular boxes denote unitary gates, representing the control subroutine steps, while diamond-shaped symbols represent non-unitary gates, used to depict the thermalization subroutine steps. Each protocol involves iterating these corresponding rounds. The explicit form of each gate is detailed in the main text.}
\label{fig_circuits}
\end{figure}

In the subsequent sections, we will delve into each protocol's structure and specific forms of control and thermalization operations examined in this work. The circuit views of all the protocols under investigation are collectively presented in\cref{fig_circuits}.

\section{Performance of HBAC protocols}\label{sec_HBAC_performance}

The standard approach for assessing the efficacy of HBAC protocols involves evaluating their cooling limits, 
namely, determining the lowest achievable temperature of the target qubit given the constraints of machine size and permitted operations. 
The cooling limit is reached asymptotically and can be determined by observing the system's long-term behavior. A more practical strategy involves closely monitoring the target qubit, with an exclusive emphasis on maximizing its cooling efficiency, while disregarding the evolution in the machine's state. Within this framework, we present the cooling limits of the different HBAC protocols using the \emph{transition matrix} formalism, which is well-suited for this analysis.
This section provides a concise overview of the transition matrix formalism. We then apply this formalism to each HBAC protocol presented in this paper to gain a unified view of their cooling behavior. Finally, in subsection (\ref{UnifiedCooling}),  we present the unified cooling limits.

\subsection{Transition matrix formalism}\label{sec_transition_matrix}


The state of a general quantum system is described by a density operator, which can be represented in vector form by choosing some orthonormal basis $\sigma_{k}$.
This is typically represented by a supervector $\dket{\rho}$, with components $\dket{\rho}_{k} = \langle \sigma_{k}, \rho \rangle$ where $\sigma_{k}$ ranges over all the basis elements.
Given a completely-positive trace-preserving (CPTP) map $\mc{E}$, if we construct a matrix $G_{\mc{E}}$ with the $i,j$-th component being $\langle \sigma_{i}, \mc{E}(\sigma_{j}) \rangle$ where $\sigma_{i}$, $\sigma_{j}$ ranges over the basis elements, then this matrix entails all the information of $\mc{E}$.
This matrix is called the transition matrix for the process $\mc{E}$, and it acts on the state $\dket{\rho}$ through matrix multiplication:
\begin{equation}\label{eqn_G_matrix}
    \dket{\mc{E}(\rho)} = G_{\mc{E}} \dket{\rho}
\end{equation}
by a simple linearity argument.

One property of $G_{\mc{E}}$ is that it is left-stochastic, which comes from the trace-preserving requirement of $\mc{E}$.
The matrix $G_{\mc{E}}$ fully describes the effect of an operation and is uniquely determined by the cooling protocol.
Note that the output of $G_{\mc{E}}$ may have a different dimension than the input state.

Knowing the exact form of $G$ for each step enables a direct simulation of the cooling process through direct matrix multiplication.
For certain HBAC protocols that consist of repeated rounds of the same operation, the computation is greatly simplified by first diagonalizing $G$ as $G = T D T^{-1}$ where $D$ is diagonal (for the cooling methods presented, their matrix $G$ is diagonalizable), then raising $D$ to the desired power $N$ where $N$ is the number of rounds. 
Taking the limit of $N \rightarrow \infty$ gives the asymptotic final state of the system, from which quantities such as the asymptotic polarization can be easily computed.
We utilize this technique to derive several important characterizations for different protocols, including the polarization evolution, the asymptotic state, as well as thermodynamic quantities like work and heat. Furthermore, this approached allowed us to unify the different HBAC protocols.


\subsection{Single shot entropy compression\\
(Algorithmic cooling without a bath)}\label{sec_single_shot}

The first approach to algorithmic cooling, given by Schulman and Vazirani~\cite{Schulman1998,Schulman1999a}, aimed to decrease the entropy in a target subsystem by implementing reversible operations that redistribute the internal entropy within the entire system. 
This initial approach focused exclusively on the use of reversible operations on the system, here called \textit{entropy compression} operation, without considering pumping entropy out into an external reservoir. 
The lack of an external reservoir makes this method a single-shot process constrained by the conservation of the eigenvalues of the global system. 

We present the maximum achievable cooling limit for single-shot reversible entropy compression on a string of $n$ identical qubits. It is important to note that Schulman and Vazirani proposed explicit protocols for implementing entropy compression that differ from the optimized strategies detailed in this section.

The system setup consists of a string of $n$ identical qubits, all initially in a product state, where each individual qubit is in the state $\rho=\begin{pmatrix}p & 0 \\ 0 & 1-p\end{pmatrix}$, with $p\geq 1/2$.
For simplicity and without limiting the generality of the approach, the first qubit of the string is designated as the target qubit to be cooled.

The total state of the system, denoted as $\rho_T=\rho^{\otimes n}$, is a diagonal matrix with $2^n$ elements, each taking the form $p^{n-i}(1-p)^i$, for $i=0,1,...,n$. To be more precise, the diagonal of this state matrix features $\binom{n}{i}=\frac{n!}{i!(n-i)!}$ entries of the element $p^{n-i}(1-p)^i$ for each $i=0,1,\ldots,n$.

To derive the maximum achievable ground state population for the target qubit, denoted by $q_{max}$, under global unitary operation on the entire system, we make use of the Schur-Horn theorem.

First, note that the ground state population of the target qubit, which is positioned first in the string, corresponds to the sum of the first half elements in the diagonal of the total system. Therefore, the goal is to enhance that specific partial sum of the diagonal entries in the updated total density matrix. Namely,
\begin{equation}
    q_{max}= \max_{U} \sum_{k=1}^{2^{n-1}} (U\rho_T U^\dagger)_{kk}
\end{equation}
where $(U\rho_T U^\dagger)_{kk}$ is the $k^{\rm th}$ diagonal element of the transformed density matrix, with the maximization performed over all unitary transformations $U$ on the total system.

The Schur–Horn theorem~\cite{schur1923uber,horn1954doubly} establishes a relationship between the diagonal elements and eigenvalues of a Hermitian matrix. According to the theorem, for a density matrix $\rho$ with N diagonal elements $\{\rho_{ii}\}_{i=1}^\textsc{n}$, the following conditions with respect to its eigenvalues $\{\lambda_i\}_{i=1}^\textsc{n}$ are always satisfied:
\begin{align}
    \sum_{i=1}^m\rho_{ii}\leq\sum_{i=1}^m \lambda_{i}^\downarrow, \quad {\rm for }\quad \forall m=1,2,...,N-1\\
    {\rm and} \quad  \sum_{i=1}^N\rho_{ii}=\sum_{i=1}^N \lambda_{i}^\downarrow
\end{align}
where $\{\lambda_i^\downarrow\}$ denotes the sequence of eigenvalues arranged in non-increasing order. Essentially, this indicates that the diagonal vector of a density matrix is always majorized by the vector of its eigenvalues.

Thus, the Schur-Horn theorem's inequalities define a reachable upper limit in terms of the eigenvalues $\{\lambda_i\}$ of $\rho_T$, as follows:
\begin{equation}
    q_{max}= \max_{U} \sum_{k=1}^{2^{n-1}} (U\rho_T U^\dagger)_{kk}=
    \sum_{i=1}^{2^{n-1}} \lambda_{i}^\downarrow.
\end{equation}
This upper bound is precisely achieved when a unitary transformation $U$ leaves the state diagonal and places the highest populations within the first half of the diagonal elements. Consequently, employing such a unitary transformation represents the optimal strategy for saturating this upper bound.

To calculate the maximum ground state population for our system, note that, due to the absence of coherences, the diagonal elements of $\rho_T$ directly represent its eigenvalues. So, let us arrange these diagonal elements in descending order; given that $p\geq 1/2$, the sequence is as follows:
\begin{equation}
p^n\geq p^{n-1}(1-p)\geq p^{n-2}(1-p)^{2}\geq ...\geq (1-p)^n 
\label{eq:order}
\end{equation}
that is, $p^{n-i}\left(1-p\right)^i\geq p^{n-j}\left(1-p\right)^j$ for all $i\leq j$.

Then, by partially summing the first half of these ordered elements, we obtain $q_{max}$ as a function of the number of qubits $n$ and the initial probability $p$, which gives the result and proof of our following theorem.

\begin{theorem} \textnormal{\textbf{Cooling limit for a single-shot reversible entropy compression on n qubits.}} The maximum achievable ground state population obtained by a unitary entropy compression on a system of n identical qubits, all initially in a product state, where each individual qubit is in the state $\rho=\begin{pmatrix}p & 0 \\ 0 & 1-p\end{pmatrix}$, with $p\geq 1/2$, is given by
\begin{equation}
\begin{aligned}
& q_{max}= 
\begin{cases}
    \displaystyle \sum^{(n-1)/2}_{i=0}\binom{n}{i}p^{n-i}\left(1-p\right)^{i}, \quad \text{for odd}\; n\\
    \\
    \displaystyle \sum^{n/2}_{i=0}\binom{n}{i}p^{n-i}\left(1-p\right)^{i}-\frac{1}{2}\binom{n}{n/2}p^{n/2}\left(1-p\right)^{n/2}\\ \text{for even } \: n
\end{cases}
\end{aligned}
\label{eq:pp_entropy_compression}
\end{equation}
This cooling limit is attainable by unitary transformations that leave the total state diagonal with the largest populations in the first half of the diagonal elements.
\end{theorem}

The $q_{max}$ corresponds to the cumulative distribution function of a binomial distribution:
\begin{equation}
    F(k,n,1-p):= \sum^k_{i=0}\binom{n}{i}p^{n-i}\left(1-p\right)^{i},
\end{equation}
and can also be represented by the regularized incomplete beta function, as follows:
\begin{equation}
\displaystyle
    F\left(k,n,1-p\right)= \left(n-k\right)\binom{n}{k} \int_0^{p}t^{n-k-1}\left(1-t\right)^k dt
\end{equation}
Note that $q_{max}(p,2j+1)=q_{max}(p,2j)$, for $j=1,2,3,...$.

\begin{corollary} In the thermodynamic limit, the maximum achievable polarization of the target qubit can be approximated to 
\begin{equation}
\displaystyle
    \epsilon_{target}=\frac{1}{\sqrt{\pi}}\int^\mu_0 e^{-t^2}dt
\end{equation}
where $\mu$ is defined as
\begin{align*}
    \mu &=\frac{n/2-n(1-p)}{\sqrt{2np(1-p)}}=\frac{n\epsilon_b}{\sqrt{2n(1-\epsilon_b^2)}},
\end{align*}
and $\epsilon_b$ is the initial polarization of the qubits, $\epsilon_b=2p-1.$
\end{corollary}

This result is derived by applying the Central Limit Theorem to the Theorem 1. This result provides a good approximation even for small values of $n$. In the regime where the polarization is small, the expression becomes $\mu\approx\sqrt{\frac{n}{2}}\epsilon_b$.


\begin{figure}[ht]
\centering
\begin{subfigure}{1\linewidth}
    \includegraphics[width=\linewidth]{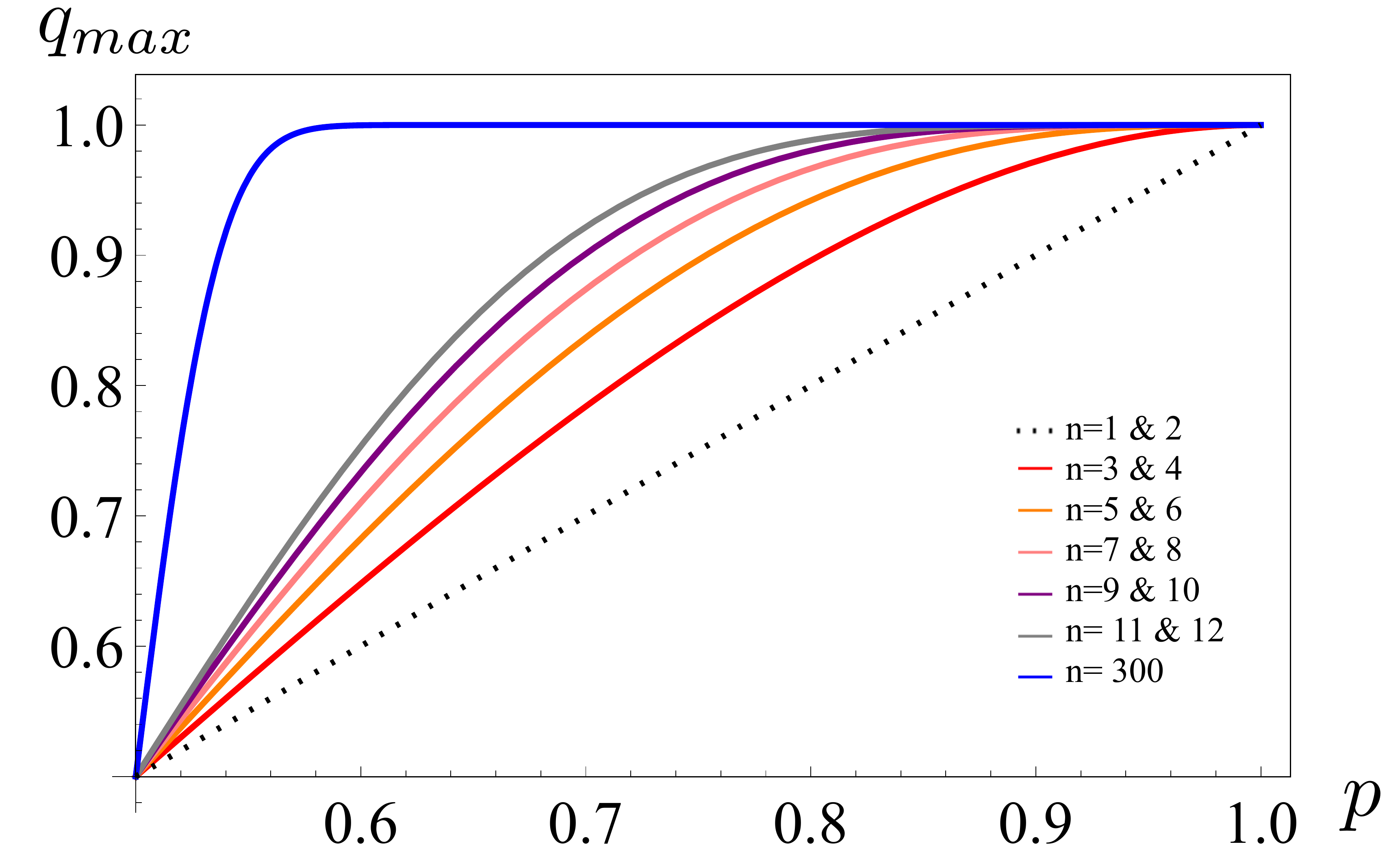}
\end{subfigure}

\caption{Maximum achievable ground state population $q_{max}$ of the target qubit after a shot of entropy compression as a function of the initial population $p$, for different number of qubits. The black dotted line represents the baseline where no improvement in population occurs. Note that configurations with odd $n$ number of qubits and the subsequent even number ($n+1$) have identical final population values.}
\label{fig_pp_entropy_compression}
\end{figure}

For qubits in a thermal state at inverse temperature $\beta_b$, and gap $\omega$, the initial population $p$ is given by 
\begin{equation}
    p=p_{b} \coloneqq \frac{1}{1+e^{-\beta_{b} \omega}}.
    \label{eq:pb_p}
\end{equation}
Following a single shot of entropy compression, the effective local temperature of the target qubit is reduced, leading to a new effective inverse temperature $\beta'$:
\begin{equation}
\beta'\omega=\log[\frac{q_{\text{max}}}{1-q_{\text{max}}}].
\end{equation}

By replacing eqs.(\ref{eq:pp_entropy_compression}), (\ref{eq:pb_p}), and $\beta_b=T_b^{-1}$, and considering the low and high temperature regimes, we obtain the following:

\begin{itemize}
      \item In the regime of low bath temperature $T_b$: the new local temperature $T'$ approaches $\frac{2}{n+1} T_b$ for odd $n$ maintaining this value for the subsequent even number $n+1$. For example, for $n=3$ and $4$: $T'\to \frac{T_b}{2}$; $n=5$ and $6$: $T'\to \frac{T_b}{3}$; and so on.
    \item In the high temperature regime, the modified local temperature $T'$ approaches $T_b / \alpha(n)$ for both odd $n$ and the subsequent even number. The function $\alpha(n)$ is approximately given by $0.886 + 0.226n - 0.006665n^2 + 0.0001n^3$, capturing the nuanced dependence of the cooling effect on the system size characterized by the number of qubits $n$.
\end{itemize}

\subsection{PPA}\label{subsec_Gmatrix_PPA}
In the PPA cooling protocol, the control subroutine features a compression unitary $U_{\textsc{ppa}}$ designed to perform a decreasing SORT operation on the diagonal elements of the input state. The sort operation is with respect to an ascending, lexicographical computational basis where $000$ comes before $001$, which comes before $010$, and so on.
Note that the arrangement of qubits plays a crucial role in this process, and the form of that $U_\textsc{ppa}$ is optimized for the scenarios where the target corresponds to the the first qubit in the system. 
After this control subroutine, the population in the $0$ subspace of the first qubit increases, implying a local entropy decrease. Since the global entropy of the system is conserved under unitary operations, this step leads to a separation of cold and warm regions within the system, where the target defines the ``cold side''.

The thermalization subroutine consists of a refresh of qubits at the warm side by rethermalizing them with the thermal bath.
In the analysis presented in this work, we focus on relaxing the two hottest qubits, which were observed to exhibit higher temperatures than the thermal bath following the execution of all control subroutines. 
We assume that the reset qubits' relaxation occurs significantly faster than the natural evolution of the other qubits in the system, so their states remain effectively unchanged during the time it takes for the reset qubits to thermalize.

Analyzing the PPA poses a challenge due to the complexity involved in specifying the explicit form of the entropy compressions, as they vary based on the system's state. Although the effect of the $U_\textsc{ppa}$ is known, its matrix representation changes with each round in systems comprising more than three qubits.
Meanwhile, the 3-qubit case, which represents the smallest scenario within the PPA framework, stands as an exception. It is characterized by the property that all its compression unitaries are identically equal to~\cite{Rodriguez-Briones2016}
\begin{equation}\label{eqn_U_PPA3_comp}
    U_{\text{PPA3}} = \begin{pmatrix}
        1 & 0 & 0 & 0 & 0 & 0 & 0 & 0\\
        0 & 1 & 0 & 0 & 0 & 0 & 0 & 0\\
        0 & 0 & 1 & 0 & 0 & 0 & 0 & 0\\
        0 & 0 & 0 & 0 & 1 & 0 & 0 & 0\\
        0 & 0 & 0 & 1 & 0 & 0 & 0 & 0\\
        0 & 0 & 0 & 0 & 0 & 1 & 0 & 0\\
        0 & 0 & 0 & 0 & 0 & 0 & 1 & 0\\
        0 & 0 & 0 & 0 & 0 & 0 & 0 & 1
    \end{pmatrix}.
\end{equation}
The thermalization of a single qubit is represented by the following reduced transition matrix:
\begin{equation}
    G_{\Gamma_{1}} = \begin{pmatrix}
        p_{b} & p_{b}\\
        1-p_{b} & 1-p_{b}
    \end{pmatrix},
\end{equation}
where $p_b$ given in \cref{eq:pb_p}. 
Then, the ground state population of the target qubit, denoted $p_{t(\ppa)}$ at the end of the $N^{\rm th}$ round is
\begin{equation}\label{eqn_PPA3_population_evo}
    p_{t(\ppa)}(N) = \frac{z}{1+z} - y^{N} (\frac{z}{1+z} - p_{t}),
\end{equation}
where $y \coloneqq  2p_{b}(1-p_{b})$ and $z \coloneqq p_{b}^{2}/(1-p_{b})^{2}$.
Then, the corresponding asymptotic state is
\begin{equation}
    \dket{\rho_{t(\ppa)}(\infty)} = \frac{1}{1+z} \begin{pmatrix}
    z \\ 1
    \end{pmatrix}.
\end{equation}
 The derivation of these formulas, which employs the transfer matrix formalism, is elaborated in Appendix~\ref{sec_appen_calc}.

For systems with more than 3 qubits, the unitary operation $U_{\textsc{ppa}n}$ varies from one round to the next. To date, an analytical expression for the unitary during a general round of PPA with
$n>3$ has not been established. Consequently, to investigate the PPA's behavior in larger systems, we will rely on numerical simulations.


\subsection{The 2-qubit NOE2-HBAC}\label{subsec_Gmatrix_NOE}
In the 2-qubit Nuclear Overhauser Effect Heat-Bath Algorithmic Cooling (NOE2-HBAC), each round consists of two distinct subroutines. The first subroutine involves randomizing the non-target qubit to a completely mixed state (CMS).
In practice, this step can be accomplished by applying random rotations to the non-target qubit and then averaging the outcomes across multiple experiments. 
The second subroutine, aimed at thermalization, employs a $\Gamma_{2}$ relaxation mechanism specifically designed to refresh the states $\ket{00}$ and $\ket{11}$, defined as follows:

\begin{definition} The $\Gamma_{2}$ relaxation process is defined by the following 6 Kraus operators:
\begin{equation}
    \begin{gathered}
    A_1 = \sqrt{p_{b2}} \ket{00}\bra{00}\\
    A_2 = \sqrt{p_{b2}} \ket{00}\bra{11}\\
    A_3 = \sqrt{1-p_{b2}} \ket{11}\bra{11}\\
    A_4 = \sqrt{1-p_{b2}} \ket{11}\bra{00}\\
    A_5 = \ket{01}\bra{01}\\
    A_6 = \ket{10}\bra{10},
    \end{gathered}
\end{equation}
where $p_{b2}$ is given by
\begin{equation}\label{eqn_defn_p2}
p_{b2} \coloneqq \frac{1}{1+e^{-2\beta_{b} \omega}},
\end{equation}
\end{definition}
\noindent in contrast to \cref{eq:pb_p}.

This $p_{b2}$ corresponds to the ground state population of a thermal qubit at bath temperature with energy gap $2\omega$, which can also be interpreted as the relative ground state population between the states $\ket{00}$ and $\ket{11}$ in thermal equilibrium, if both qubits have an energy gap of $\omega$.
In comparison, the $\Gamma_{1}$ thermalization achieves a thermal equilibrium between the states $\ket{0}$ and $\ket{1}$ for a single qubit.

The operation $\Gamma_{2}$ is summarized by the following reduced transition matrix:
\begin{equation}
    G_{\Gamma_{2}} = \begin{pmatrix}
        p_{b2} & 0 & 0 & p_{b2}\\
        0 & 1 & 0 & 0\\
        0 & 0 & 1 & 0\\
        1-p_{b2} & 0 & 0 & 1-p_{b2}
    \end{pmatrix}.
\end{equation}
This map effectively dephases the input state by eliminating the off-diagonal elements.
Furthermore, the $\Gamma_{2}$ preserves the Gibbs state $\rho_{t,m}^{th}(\beta_{b},H_{t,m})$ and acts non-trivially on only two energy levels, making it fall into the category of 2-level Gibbs stochastic matrices.
It is known that every 2-level Gibbs stochastic transition matrix corresponds to a thermal operation~\cite{Lostaglio2018}.
Therefore, it falls under the thermalization subroutine category by our definition.
%

The ground state population of the target qubit after $N$ rounds is
\begin{equation}
    p_{t(\noe 2)}(N) = \frac{x}{1+x} - 2^{-N} (\frac{x}{1+x} - p_{t})
\end{equation}
where $\displaystyle x=\frac{p_{b2}}{1-p_{b2}}$.
The corresponding asymptotic state is
\begin{equation}
    \dket{\rho_{t(\noe 2)}(\infty)} = \frac{1}{1+x} \begin{pmatrix}
    x \\ 1
    \end{pmatrix} = \begin{pmatrix}
        p_{b2} \\ 1-p_{b2}
    \end{pmatrix}.
\end{equation}
See Appendix~\ref{sec_appen_calc} for detailed derivation.

\subsection{The 2-qubit SR2-HBAC}\label{subsec_Gmatrix_SR2}
In the 2-qubit State-Relaxation Heat-Bath Algorithmic Cooling (SR2-HBAC)~\cite{Rodriguez-Briones2017}, the control subroutine of a round is to apply a Pauli $\sigma_{x}$ gate on the non-target qubit.
The thermalization subroutine is implementing the $\Gamma_{2}$ relaxation on both qubits, followed by a $\Gamma_{1}$ relaxation in the non-target qubit.

The ground state population of the target qubit after $N$ rounds of the SR2-HBAC is
\begin{equation}
    p_{t(\srg)}(N) = \frac{w}{w+v} - u^{N}(\frac{w}{w+v} - p_{t}),
\end{equation}
where $w \coloneqq p_{b2} p_{b}$, $v = 1-p_{b2} -p_{b} + w$, and $u = 1-v-w$.
The corresponding asymptotic state is
\begin{equation}
    \dket{\rho_{t(\srg)}(\infty)} = \frac{1}{w+v} \begin{pmatrix}
    w \\ v
    \end{pmatrix}.
\end{equation}
See Appendix~\ref{sec_appen_calc} for detailed derivation.

\subsection{The 1-qubit xHBAC1} \label{subsec_Gmatrix_xHBAC}
The rounds of the 1-qubit xHBAC1 protocol comprises two main operations. Initially, a 1-qubit Pauli $\sigma_{x}$ gate, followed by a thermalization given by a $\beta$-swap operation, defined in references~\cite{Lostaglio2018,Alhambra2019} as the dephasing operation with the following Kraus operators:
\begin{equation}
\begin{gathered}
    A_{1} = \ketbra{0}{1}\\
    A_{2} = \sqrt{e^{-\beta_{b} \omega}} \ketbra{1}{0}\\
    A_{3} = \sqrt{1 - e^{-\beta_{b} \omega}} \ketbra{0}{0}.
\end{gathered}
\end{equation}

The ground state population of the target qubit after $N$ rounds of 1-qubit xHBAC cooling is
\begin{equation}
    p_{t(\xhbac)}(N) = 1- s^{N}(1-p_{t}),
\end{equation}
where $s=e^{-\beta_{b}\omega}$.
The corresponding asymptotic state is given by
\begin{equation}
    \dket{\rho_{t(\xhbac)}(\infty)} = \begin{pmatrix}
    1 \\ 0
    \end{pmatrix},
\end{equation}
i.e., the pure state $\ket{0}$. See Appendix~\ref{sec_appen_calc} for detailed derivation.

\subsection{Unified cooling limits\label{UnifiedCooling}}


The asymptotic polarization for the different algorithmic cooling protocols discussed here is captured by the general expression:
%
\begin{align}
\displaystyle
    \epsilon_{\infty}\left(\epsilon_b,\alpha\right):= \frac{\left(1+\epsilon_b\right)^{\alpha}-\left(1-\epsilon_b\right)^{\alpha}}
    {\left(1+\epsilon_b\right)^{\alpha}+\left(1-\epsilon_b\right)^{\alpha}}\\
    = \tanh{\left[\alpha * {\rm arctanh}{\left(\epsilon_b\right)}\right]}.
\end{align}
where $\alpha$ depends on the implemented protocol, as shown below. Equivalently, since $1+\epsilon= 2p$, and $1-\epsilon = 2\delta$, in the cooling limit, the ground state population of the target qubit corresponds to
\begin{equation}
p_{t,\infty}(p_{b},\alpha) = \frac{p_{b}^{\alpha}}{p_{b}^{\alpha}+\delta_{b}^{\alpha}}    
\end{equation}

The temperature of the target qubit in the asymptotic limit is given by
\begin{equation}
    T_{t,\infty}=\frac{T_b}{\alpha},
\end{equation}
where $T_b$ is the temperature of the thermal bath.

The $\alpha$ value for the various protocols are specified as follows:

\begin{itemize}
   \item \textbf{Single-shot reversible entropy compression: } In the low temperature regime,
   $\alpha = \lceil n/2 \rceil$ for $n$ qubits, when applying the optimal entropy compression (sorting the diagonal elements of the total density matrix in descending order.) 
   
    \item \textbf{PPA:} $\alpha=2^{n-2}$, for a string of $n$ qubits, with either one or two reset qubit(s).
    In general, $\alpha=md$, when implemented on a system composed of a target qubit, an auxiliary system of dimension $d$, and $m$ reset qubits. For instance, $\alpha=m2^{n'}$ when using an auxiliary system of $n'$ qubits.
      

    %
    %
    \item \textbf{NOE2:} $\alpha=2$, for two qubits. Note that the cooling limit of the NOE protocol on 2 qubits is the same as the cooling limit of the PPA-HBAC on 3 qubits.
        %

\item \textbf{SR$n$:} $\alpha=2^n-1$, for a string of $n$ qubits.
%

        
    %
    \item \textbf{xHBAC1:} $\alpha=\infty$, i.e. $\epsilon_\infty^{\rm xHBAC1}\to 1 $, $\forall  \epsilon_b$
    %
        %
\end{itemize}


\subsection{Unified polarization evolution}\label{sec_unified_evolution}


 Under the implementation of the different algorithmic cooling protocols, the polarization of the target qubit is enhanced exponentially with the number of rounds at different convergence speeds.
The general expression of the polarization as a function of the number of rounds $k$ and convergence rate $r$ is given by%
\begin{equation}
    \epsilon_k\left(r,\epsilon_\infty,\epsilon_b\right)=\epsilon_\infty -r^k\left(\epsilon_\infty-\epsilon_b\right),
\end{equation}
where $\epsilon_b$ is the polarization of the heat-bath and the parameters $r<1$, and $\epsilon_\infty$ depend on the protocol used, as follows:


\begin{itemize}
    \item \textbf{PPA:} 
          $\displaystyle r=\frac{1-\epsilon_b^2}{2}$, with  
          $\displaystyle \epsilon_\infty=\frac{2\epsilon_b}{1+\epsilon_b^2}$, for three qubits.
    %
    %
    
    \item \textbf{NOE2:}
    $r=1/2$ with $\displaystyle\epsilon_\infty=\frac{2\epsilon_b}{1+\epsilon_b^2}$, for two qubits.
    


    %
    %
        \item 
        \textbf{ SR2:} 
        $\displaystyle r=\frac{1}{2}\left(\frac{1-\epsilon_b^2}{1+\epsilon_b^2}\right)$ with\\ $\displaystyle \epsilon_\infty=\tanh{\left[3 {\rm arctanh}{\left(\epsilon_b\right)}\right]}=\epsilon_b\frac{3+\epsilon_b^2}{1+3\epsilon_b^2}$, for 2 qubits.

        \item \textbf{xHBAC1:}
        $\displaystyle r=e^{-\beta_{b} \omega}=e^{-2 {\rm arctanh} (\epsilon_b)}$ and $\epsilon_\infty=1$.


\end{itemize}

\subsection{A note on the notions of resources}
Before we delve further, it's pertinent to discuss the different notions of ``required resources'' for the AC protocols.
Although our forthcoming discussion will primarily 
focused on the thermodynamic cost, specifically the work cost associated with these protocols, there are other notions of cost discussed in the literature. The notion of control complexity and time cost have been proposed in reference~\cite{Taranto2023}, which applies to the protocols reviewed here as well.
Trade-offs are inherent between control complexity, cooling power, and the required system size. Specifically, for the most basic form of the PPA, which uses a simple full thermalization interaction with the thermal bath, requires at least three qubits to get an improvement in the target qubit.
Conversely, it is possible to cool down a target qubit from a two qubit system by utilizing the ``state reset'' interaction with the bath, which is a more detailed evolution with higher control complexity.
Finally, using xHBAC  protocol presents the theoretical possibility of approaching zero temperature with merely a single qubit; however it requires detailed, controlled interaction between the qubit and the bath, which has a diverging control complexity.
A quantitative link between these different notions of resources, particularly in scenarios with finite resources, holds significant promise for advancing our ability to control quantum systems and leverage their capabilities to enhance AC techniques.



\section{Cooling efficiency of HBAC protocols}\label{sec_HBAC_efficiency}
We are now ready to delve into the thermodynamic efficiencies of various AC protocols.
The significance of efficiency analyses spans both theoretical frameworks and practical applications. Theoretically, these analyses establish the upper bounds of what efficiencies are attainable, serving as benchmarks for the capabilities of the different AC protocols. Practically, they highlight potential areas where optimization is possible, guiding improvements in protocol design and implementation.
Traditionally, the development of quantum cooling protocols has focused primarily on enhancing the cooling limit, with little attention to the cost or efficiency of these protocols. Specifically, the main goal has been to achieve lower temperatures or higher polarization in quantum systems without thoroughly evaluating the resources required or considering the implications and practicality of implementing these protocols.
However, as quantum processors continue to grow in size and complexity, efficiency considerations are gaining prominence. The scalability of quantum technologies depends not just on their theoretical capabilities but also on their operational efficiency and resource demands. As we advance, the focus is likely to shift towards optimizing these cooling protocols to be more resource-efficient, ensuring that they can be feasibly integrated into larger quantum systems. This shift towards efficiency will be crucial in transitioning quantum cooling protocols from laboratory curiosities to foundational tools in the practical application of quantum technologies.

In the following sections, we analyze two different notions of efficiencies in quantum thermodynamics, and conduct a comparative analysis of various AC protocols based on these efficiency metrics.
Following this comparative review, we will introduce several modified AC protocols that demonstrate improved energy efficiency compared to those previously discussed.

A note on notation: throughout this section, which often deals with variations such as changes in a system's energy, we will consistently use the Greek letter $\Delta$ to denote the net change of a state function for a process. Our convention for the net change is ``final - initial'', meaning $\Delta A = A_{f} - A_{i}$, where $A_{f}$ is the final value of some quantity $A$ after the process, and $A_{i}$ is the initial value of $A$ before the process.

\subsection{Work cost and the coefficient of performance}\label{subsec_COP} 
The first and foremost metric of interest in our discussion is the minimum amount of work input required to implement the AC protocols, drawing inspiration from the classical coefficient of performance (CoP) concept.
 A classical cooling engine operates between a cold bath and a hot bath, pumping energy out of the cold bath and transferring it to the hot bath.
For any operation performed by the engine, the CoP provides a measure of how effective it is in removing energy from the cold bath, giving the ratio of the desired effect (the energy removed from the target) to the required input (work input to extract that energy). Specifically, the CoP, denoted by K, is given by
\begin{equation}\label{eqn_defn_COP}
    K \coloneqq \frac{-\Delta E_{t}}{W},
\end{equation}
where $-\Delta E_{t}$ is the total amount of energy removed from the target during the operation, and $W$ is the work input for that process.
As the definition of ``operation'' in the above varies, one can study the CoP of a single step during a protocol, or that of a complete cooling process.

We can analogously have a CoP for a quantum cooling procedure.
The HBAC protocols considered here aim to cool down a finite quantum system instead of an infinite cold bath.
In this context, the CoP given by K in \cref{eqn_defn_COP}, will have $-\Delta E_{t}$ denoting the energy decrease in the target system (which can now be interpreted as a finite cold bath), and $W$ denoting the minimum work input required.
We will restrict to processes with $-\Delta E_{t} > 0$, indicating a net decrease in energy within the target system, which is essential for effective cooling. Specifically, within HBAC context, two processes are particularly relevant: a single round, which examines the energy change in the target system after one cycle of the HBAC protocol, and the cumulative effect of the cooling procedure up to a given round, assessing the protocol's efficiency over multiple rounds
We will denote the CoP for the $N$-th round by a lower-case $k(N)$, and the cumulative CoP \emph{up to} the $N$-th round by an upper-case $K(N)$.

The next question is how $k$ or $K$ can be computed given a protocol.
It is easy to recognize that since energy is a state function, the numerator is always given by
\begin{equation}\label{eqn_defn_dE_control_1}
    - \Delta E_{t} = E_{t,i} - E_{t, f}.
\end{equation}
By further assuming that the Hamiltonians $H_{t},\ H_{m}$ both return to their initial values at the end of the unitary, the average energy changes can be simplified to
\begin{equation}\label{eqn_defn_dE_control_2}
    - \Delta E_{t} = \Tr[H_{t} (\rho_{t,i} - \rho_{t,f})],
\end{equation}
where the subscripts $i$ and $f$ stand for initial and final states of the target.

On the other hand, the definition for $W$ is slightly more subtle.
An infinitesimal change in the internal energy can be divided into two components, which may be identified as infinitesimal (average) heat and work as~\cite{Anders2013}
\begin{equation}
    dU = d \Tr[\rho H] = \Tr[d\rho\ H] + \Tr[\rho\ dH] \coloneqq \delta Q + \delta W,
\end{equation}
where $\delta$ is used to denote path-dependent infinitesimal quantities.
For a process where the evolution $\rho(t)$ and $H(t)$ from time $0$ to $\tau$ is known, the mean heat and work during the process can be calculated as
\begin{equation}
    \begin{gathered}
        Q \coloneqq \int_{0}^{\tau} dt \Tr[\dot{\rho}(t) H(t)]\\
        W \coloneqq \int_{0}^{\tau} dt \Tr[\rho(t) \dot{H}(t)]
    \end{gathered}
\end{equation}
with the first law being $\displaystyle \Delta U = \int dU = Q + W$.

In general, determining $Q$ and $W$ for a general quantum process can be complex.
However, in our case, the analysis is simplified as it only needs to be performed for coherent control AC protocols.
First, for the control subroutine, the target and machine undergo a unitary evolution, so the joint system evolves under the Schrodinger equation, $\dot{\rho}_{t,m}(t) = -\frac{i}{\hbar} [H_{t,m}(t),\rho_{t,m}(t)]$.
This implies that $Q$ vanishes:
\begin{equation}
    Q = -\frac{i}{\hbar} \int_{0}^{\tau} dt \Tr[[H_{t,m}(t),\rho_{t,m}(t)] H_{t,m}(t)] = 0
\end{equation}
since trace is invariant under cyclic permutation.
Therefore, for the control subroutine, the amount of work input is $W = \Tr[\rho(\tau) H(\tau) - \rho(0) H(0)]$, i.e., the total energy change of the full system correspond to work input.
This agrees with the intuition that, since the system is isolated from the bath in the control step, the amount of heat released into the bath is $0$.
We thus have the following expression for the control subroutine:
\begin{equation}
    W_{\text{c}} = \Delta E_{t, \text{c}} + \Delta E_{m, \text{c}}
\end{equation}
where the subscript c stands for control.

For the second part of the round, namely the thermalization subroutine, we have established in \cref{subsec_coherent_control} that the transformation on the system could be achieved by 
some energy-preserving unitary $V_{[t,m,b]}$ between the system and the bath.
Thus, the thermalization step could be performed without work input in principle.
Of course, it is possible to construct another unitary $V_{[t,m,b]}'$ that achieves the same transformation on the system, but with a nonzero work input.
Therefore, the exact form of the thermalization unitary $V_{[t,m,b]}$ is non-unique and may be unknown in practice.
So, instead, we will consider the lower bound situation for the work calculation, where the thermalization on the system is achieved by an energy-preserving unitary, so that the work input is indeed zero for the thermalization part,
\begin{equation}
    W_{\text{th}} = 0,
\end{equation}
where the subscript th stands for thermalization.
This gives a recipe to compute the total lower bound work for a coherent cooling procedure consisting of any combination of control and thermalization subroutines as
\begin{equation}
    W = \sum_{\alpha} W_{\text{c}, \alpha},
\end{equation}
where $\alpha$ denotes different control steps of the protocol.
As a result, the CoP for coherent control cooling protocols can be expressed purely in terms of (changes in) state functions.
The CoP computed under this assumption will serve as an upper bound of the actual CoP in practice, due to the aforementioned reasons.



\subsection{Comparing the CoP of the cooling protocols}
We utilize the transition matrix formalism introduced in \cref{sec_transition_matrix} to systematically compute the efficiencies for different protocols.  We monitor the evolution of both the target system and the machine---thus, involving matrices of larger size.
Since most results can be derived through similar direct calculations, we present the final results here, while the detailed derivations are provided in \cref{sec_appen_calc}.

\begin{enumerate}
\item \textbf{PPA}

In PPA3, the first step involves a unitary entropy compression on the 3 qubits of the system, followed by a complete thermalization of both reset qubits with a bath at temperature $\beta_{b}$.
We will consider the scenario where all 3 qubits have the same energy gap $\omega$.
The energy change during the first step, the control subroutine, corresponds to the total work input for each round, which can be calculated as
\begin{equation}
    w_{\ppa}(N) = \omega (p_{b}-\frac{1}{2}) y^{N},\  N \geq 1,
\end{equation}
for the $N$th round, and $y \coloneqq  2p_{b}(1-p_{b})$, from \cref{eqn_PPA3_population_evo}.
The energy decrease in the target, denoted by $-\Delta e_{t}$, is given by
\begin{equation}
    -\Delta e_{t(\ppa)}(N) = \omega (p_{b}-\frac{1}{2}) y^{N},\  N \geq 1,
\end{equation}
for the $N$th round. 

Since $ w_{\ppa}(N)=-\Delta e_{t(\ppa)}(N)$, for any $N$th round in the three-qubit case,  the CoP is given by
\begin{equation}
    k_{\ppa}(N) = 1,\ N\geq 1,
\end{equation}
and the cumulative CoP by
\begin{equation}
    K_{\ppa}(N) = 1,\ N\geq 1.
\end{equation}

These results are consistent with the analysis and experimental findings documented in the work by Soldati et al.~\cite{Soldati2022}.
For PPA on more than 3 qubits, we use numerical simulation to compute the cumulative CoP at the end of each round.

\item \textbf{NOE2}

In NOE2-HBAC the first step of a round involves a driven operation that brings the non-target qubit to the CMS state, followed by a 2-qubit $\Gamma_{2}$ relaxation.
We will assume that the qubits of the system have the same energy gap $\omega$.
The energy change during the CMS operation is the work input. This is achieved by applying a random rotation unitary to qubit 2 and then averaging over the results in practice.
The work input per round and up to $N$ rounds are calculated, respectively, as
\begin{equation}
    w_{\noe}(N) = \begin{cases}
         \omega (p_{b} - \frac{1}{2}), & N=1 \\
         \omega (p_{b2} - p_{b}) 2^{1-N}, & N>1
    \end{cases}
\end{equation}
and
\begin{align}
    W_{\noe}(N) &= \sum_{i=1}^{N} w_{\noe}(i) \cr 
    &= \omega ((p_{b2} - \frac{1}{2}) - 2^{1-N} (p_{b2} - p_{b})),
\end{align}
where $p_{b2}$ and $p_{b}$ are both functions of the bath temperature, given by \cref{eq:pb_p,eqn_defn_p2}.

The energy decrease in the target qubit during the $N$-th round and up to the $N$-th round are calculated, respectively, as 
\begin{equation}
\begin{aligned}
    -\Delta e_{t(\noe)}(N) &= \omega (p_{b2} - p_{b}) 2^{-N},
\end{aligned}
\end{equation}
for all $N \geq 1$, and
\begin{equation}
    -\Delta E_{t(\noe)}(N) = (p_{b2}-p_{b}) (1-2^{-N}) \omega.
\end{equation}
Therefore, the CoP for the $N$-th round is
\begin{equation}
    k_{\noe}(N) = \begin{cases}
        \frac{p_{b2}-p_{b}}{2 p_{b}-1}, & N=1 \\
        \frac{1}{2}, & N>1.
    \end{cases}
\end{equation}

The cumulative CoP up to round $N$ is given by
\begin{equation}
    K_{\noe}(N) = \frac{(p_{b2}-p_{b}) (1-2^{-N})}{(p_{b2} - \frac{1}{2}) - 2^{1-N} (p_{b2} - p_{b})}.
\end{equation}

\item \textbf{SR2}

In the SR2-HBAC protocol, each round begins with the application of a $\sigma_{x}$ gate on the non-target qubit; followed by a $\Gamma_{2}$ relaxation process on both qubits; finishing with a complete thermalization of the non-target qubit with the bath.
We will assume that both qubits have the same energy gap $\omega$.
The energy change during the first step of the round, the $\sigma_{x}$ gate, corresponds to the work input.
The work input value is given by
\begin{equation}
    w_{\text{SR2}}(N) = \omega (2p_{b}-1),
\end{equation}
for all $N \geq 1$.
The cumulative work up to round $N$ is given by
\begin{equation}
    W_{\text{SR2}}(N) = N \omega (2p_{b}-1)
\end{equation}
The energy decrease for the target during the $N$-th round is
\begin{equation}
    -\Delta e_{t({\text{SR2}})}(N) = \omega p_{b} (1 - p_{b}) (2p_{b2}-1) u^{N-1}
\end{equation}
for all $N \geq 1$.
The total energy decrease in the target is 
\begin{equation}
    -\Delta E_{t({\text{SR2}})}(N) = \omega p_{b} (1 - p_{b}) (2p_{b2}-1) \frac{1-u^{N}}{1-u}
\end{equation}

Therefore the CoP of the $N$-th round is
\begin{equation}
    k_{{\text{SR2}}}(N) = \frac{p_{b} (1 - p_{b}) (2p_{b2}-1)}{2p_{b}-1} (p_{b2}+p_{b}-2p_{b2}p_{b})^{N-1}
\end{equation}
for all $N\geq 1$.
The cumulative CoP up to round $N$ is
\begin{equation}
    K_{\srg}(N)= \frac{p_{b} (1 - p_{b}) (2p_{b2}-1)}{2p_{b}-1} \frac{1-u^{N}}{N(1-u)}.
\end{equation}

\item \textbf{xHBAC1}

In 1-qubit xHBAC1, in each round, one first applies a $\sigma_{x}$ gate, followed by a $\beta$-swap relaxation $\Lambda_{\beta}$.
We assume that the qubit has an energy gap of $\omega$.
The work input corresponds to the energy change during the $\sigma_{x}$ gate.
For the $N$-th round, the work input is given by
\begin{equation}
    w_{\xhbac}(N) = \omega(1-2s^{N-1}(1-p_{b}))
\end{equation}
and the total work input is
\begin{equation}
    W_{\xhbac}(N) = \omega \left(n-2(1-p_{b}) \frac{1-s^{N}}{1-s} \right).
\end{equation}
The energy decrease for the target qubit during the $N$-th round is
\begin{equation}
    -\Delta e_{t({\xhbac})}(N) = \omega (1 - p_{b})(1-s) s^{N-1},
\end{equation}
for all $N \geq 1$.
The total energy decrease up to round $N$ is
\begin{equation}
    -\Delta E_{t({\xhbac})}(N) = \omega(1-p_{b})(1-s^{N}).
\end{equation}
The CoP of the $N$-th round is
\begin{equation}
    k_{{\xhbac}}(N) = \frac{(1 - p_{b})(1-s)}{1-2s^{N-1}(1-p_{b})} s^{N-1},
\end{equation}
for all $N\geq 1$, where $s=e^{-\omega \beta_{b}}$.

The cumulative CoP up to round $N$ is given by
\begin{equation}
    K_{{\xhbac}}(N) = \frac{(1-p_{b})(1-s)(1-s^{N})}{n(1-s)-2(1-p_{b}) (1-s^{N})}
\end{equation}
\end{enumerate}

\begin{figure}
\centering
\begin{subfigure}{0.96\linewidth}
    \includegraphics[width=\linewidth]{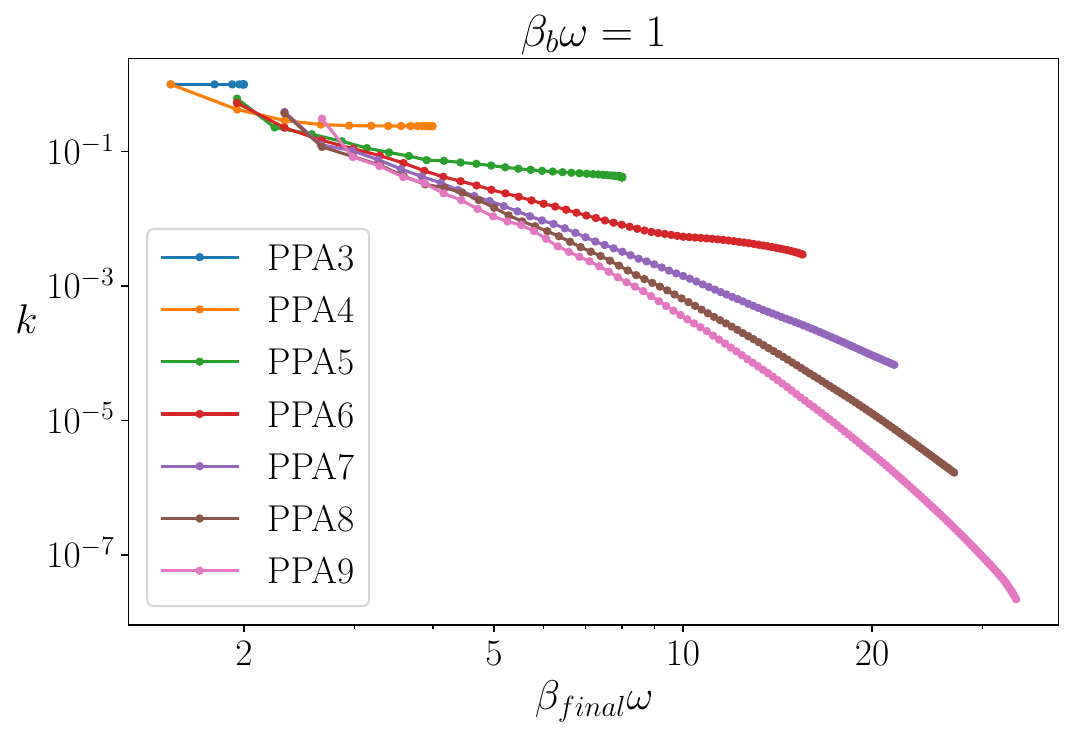}
\end{subfigure}
\caption{A plot of per-round CoP, $k_{\ppa}$, plotted against the final dimensionless inverse temperature $\beta_{\text{final}} \omega$ for PPA$n$ protocols on $n$ qubits.}
\label{fig_k_PPA}
\end{figure}

We first discuss the per-round CoP, $k$, of these protocols.
The PPA3 exhibits the simplest behavior with a constant $k_{\ppa}=1$.
The NOE has an almost constant behavior, with $k_{\noe}=1/2$ for all rounds except the first one.
Protocols like these are considered to be the most energetically effective, since as the target is cooled, the work input required at each round also decreases proportionally with the decrease in target energy per round.
Meanwhile, the SR2 and xHBAC1 both have $k$'s that decreases exponentially with the round $N$.
This means that as the qubit is cooled, the amount of energy extracted from the target becomes exponentially small compared to the work input for that round.
Therefore, protocols like these become more energetically inefficient for deeper cooling.

Next we turn to PPAn with $n>3$, where we numerically study their efficiency.
This is plotted in \cref{fig_k_PPA}.
Each point shows the per-round CoP on the y-axis, versus the dimensionless final temperature ($\beta_{\text{final}} \omega$) of the target qubit on the x-axis, so that we go from left to right as the target cools down.
The first point (on the leftmost) in each curve represents the cumulative CoP after the first round of cooling, and one follows the line towards the right as cooling proceeds further.
Note that except the PPA3 which has a constant $k$, all larger instances of PPA have decreasing $k$'s as cooling progresses.
This shows that larger PPA protocols also becomes inefficient at extracting energy from the target qubit as it is being cooled down.

\begin{figure}[ht!]
\centering
\begin{subfigure}{0.95\linewidth}
    \includegraphics[width=\linewidth]{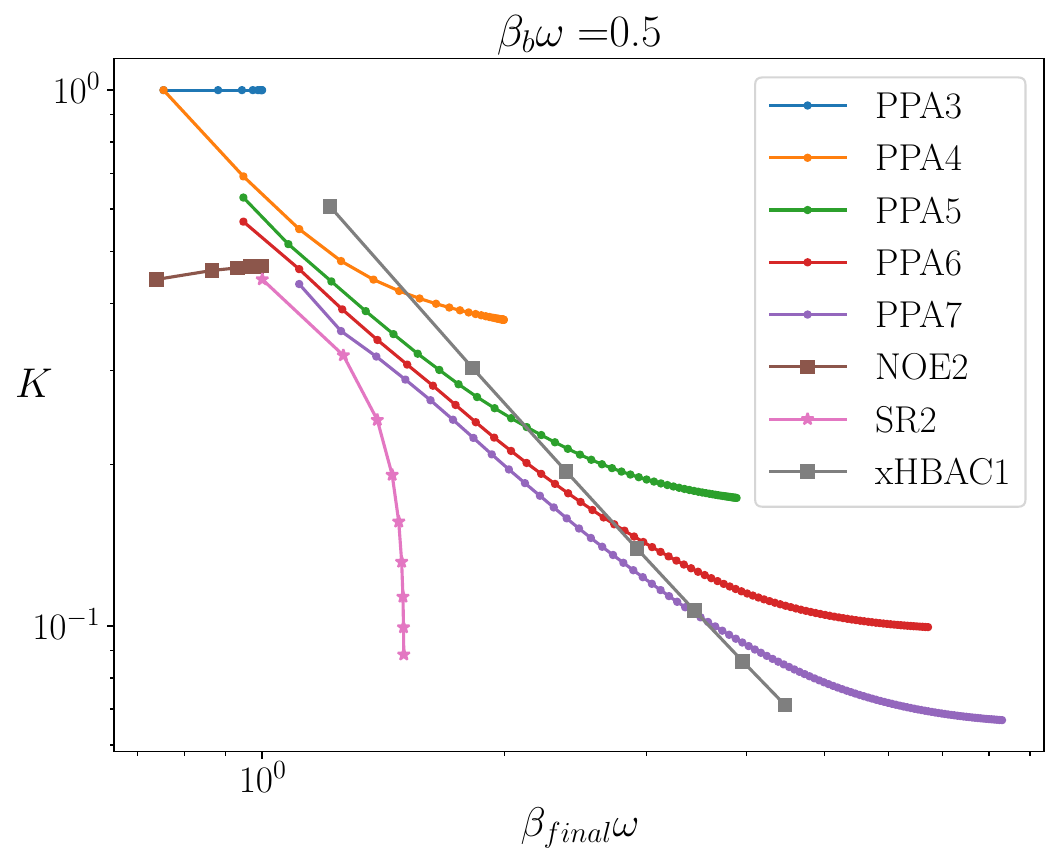}
\end{subfigure}
\hfill
\begin{subfigure}{0.95\linewidth}
    \includegraphics[width=\linewidth]{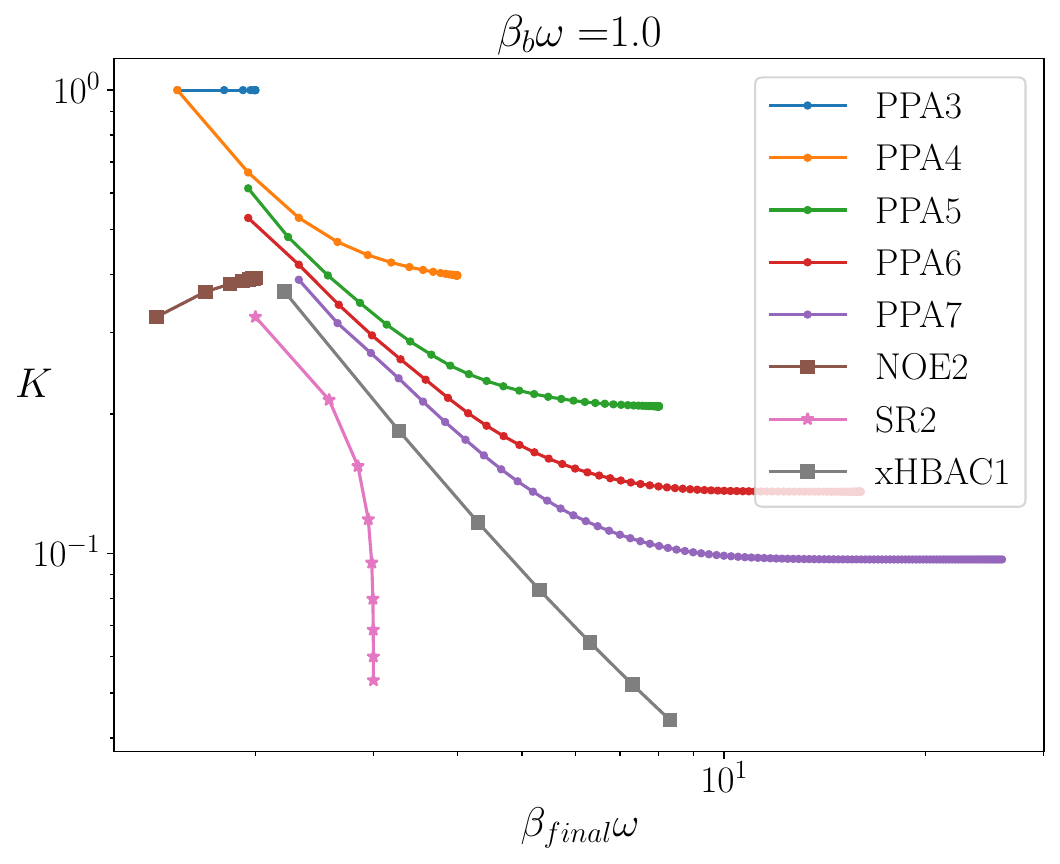}
\end{subfigure}
\hfill
\begin{subfigure}{0.95\linewidth}
    \includegraphics[width=\linewidth]{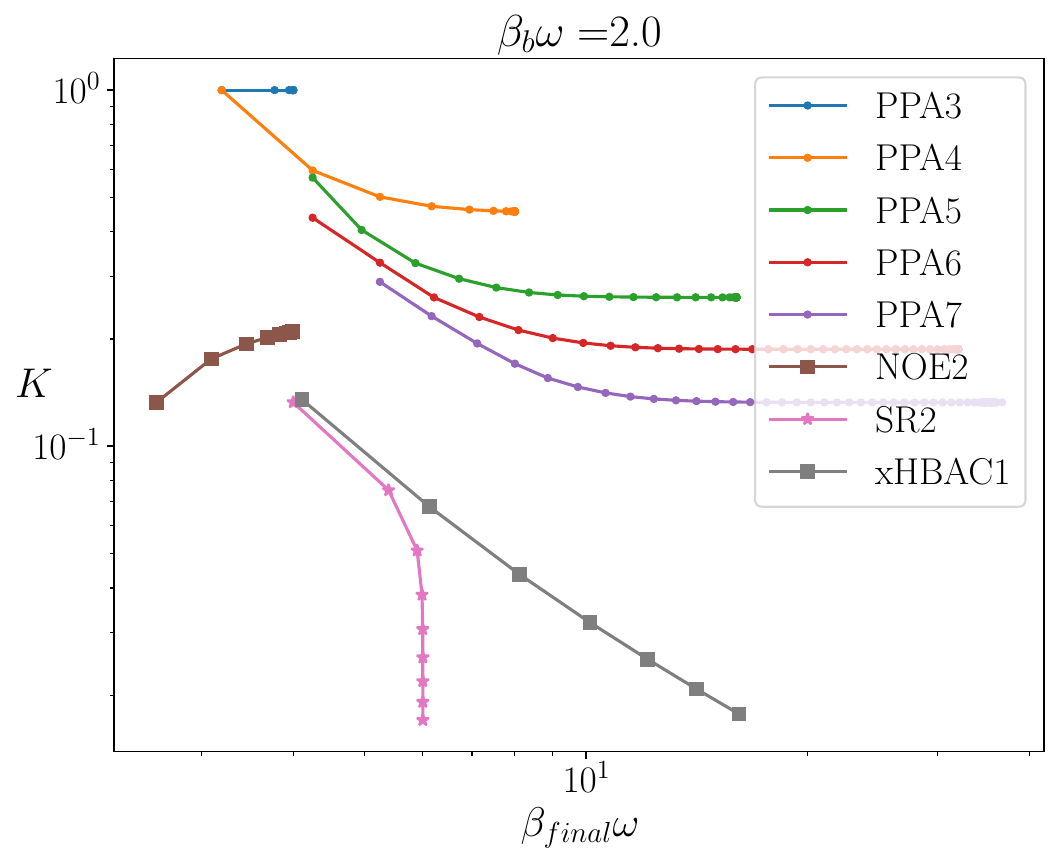}
\end{subfigure}
\caption{Comparison of cumulative CoP between HBAC protocols under high ($\beta_{b} \omega = 0.5$), intermediate ($\beta_{b} \omega = 1$), and low ($\beta_{b} \omega = 2$) initial bath temperatures, plotted against the final dimensionless temperature $\beta_{\text{final}} \omega$ of the target qubit.}
\label{fig_COP}
\end{figure}

Although the per-round CoP decreases for most of the protocols considered here, the magnitude of the work input also decreases as cooling proceeds (while the heat removed from target decreases even faster), and contributes less and less to the total amount of work inputted.
Therefore, the more practically meaningful figure of merit for determining the total energy efficiency is $K$, the cumulative CoP.
We plot the cumulative CoP for selected protocols in \cref{fig_COP}.
From top to bottom are the results when starting from a high ($\beta_{b} \omega = 0.5$), intermediate ($\beta_{b} \omega = 1$), and low ($\beta_{b} \omega = 2$) bath temperature (scaled by the qubit's energy gap $\omega$), where $\beta_{b}$ is the bath inverse temperature and $\omega$ represents the qubit energy gap.
Recall from \cref{sec_HBAC_performance} that the xHBAC has a cooling limit of $0$ (the lowest possible), whereas all other protocols under consideration here have finite cooling limits.
Here, a higher CoP implies better energy efficiency in extracting a certain total amount of heat from the target qubit.

Several important observations are listed as follows:
\begin{enumerate}
    \item The CoP for both the xHBAC1 and the SR2 goes towards 0 as cooling proceeds, whereas the other protocols have a finite CoP as we approach equilibrium;
    \item Both tending towards $K=0$ as the number of cooling rounds increase, the xHBAC1 is capable of cooling to $0$ temperature whereas SR2 has a finite equilibrium temperature;
    \item The CoP of PPA protocols plateau as we approach equilibrium, and PPA protocols with more qubits are capable of cooling to lower temperatures at the cost of having a lower final cumulative CoP;
    \item Both the PPA3 and NOE2 achieve the same final target temperature, but the PPA3 has a higher CoP than the NOE2;
    \item The relative magnitude of $K$ between different protocols vary as the bath temperature changes.
\end{enumerate}

The diverging behavior of both the xHBAC1 and the SR2 can be easily understood from the construction of their protocols, which involves applying $\sigma_{x}$ gates at every round of cooling.
Meanwhile, the optimized relaxation scheme of xHBAC1 allows it to achieve final temperature 0 exponentially fast, which SR2 cannot do.
On the other hand, the PPA appears to be a milder family of protocols, which cools to a finite temperature at a finite energy cost.
We can also see that the constant $K=1$ performance of PPA3 is a special case rather than a general behavior for all PPA protocols.

The temperature dependence of HBAC is also interesting.
It appears that as temperature lowers (going from top to bottom in \cref{fig_COP}), the PPA becomes more advantageous energetically compared to the other protocols.
Conversely, protocols like the xHBAC1 becomes more favorable when the system starts at higher initial temperatures, because it may cool to the same temperature as the PPA's in less number of steps and with less total energy input, under certain regimes.
Indeed, comparing the energy cost of protocols in \cref{fig_COP} can be done by drawing a vertical line at a certain final target temperature which we wish to achieve.
For example, we can see that at the final temperature achieved by the first round of xHBAC1 cooling, the point representing xHBAC1 lies above the ones for the PPA protocols which are capable of cooling to a similar temperature.
The second round of xHBAC1 lies below PPA4, but above PPA5 and PPA6.
On the other hand, the line for xHBAC1 lies below all PPA protocols under consideration here, for lower initial temperatures $\beta_{b} \omega = 1$ and $\beta_{b} \omega = 2$.
In summary, selecting the best protocol in practice requires one take into consideration both time and energy cost.

\begin{figure}[ht]
\centering
\begin{subfigure}{0.96\linewidth}
    \includegraphics[width=\linewidth]{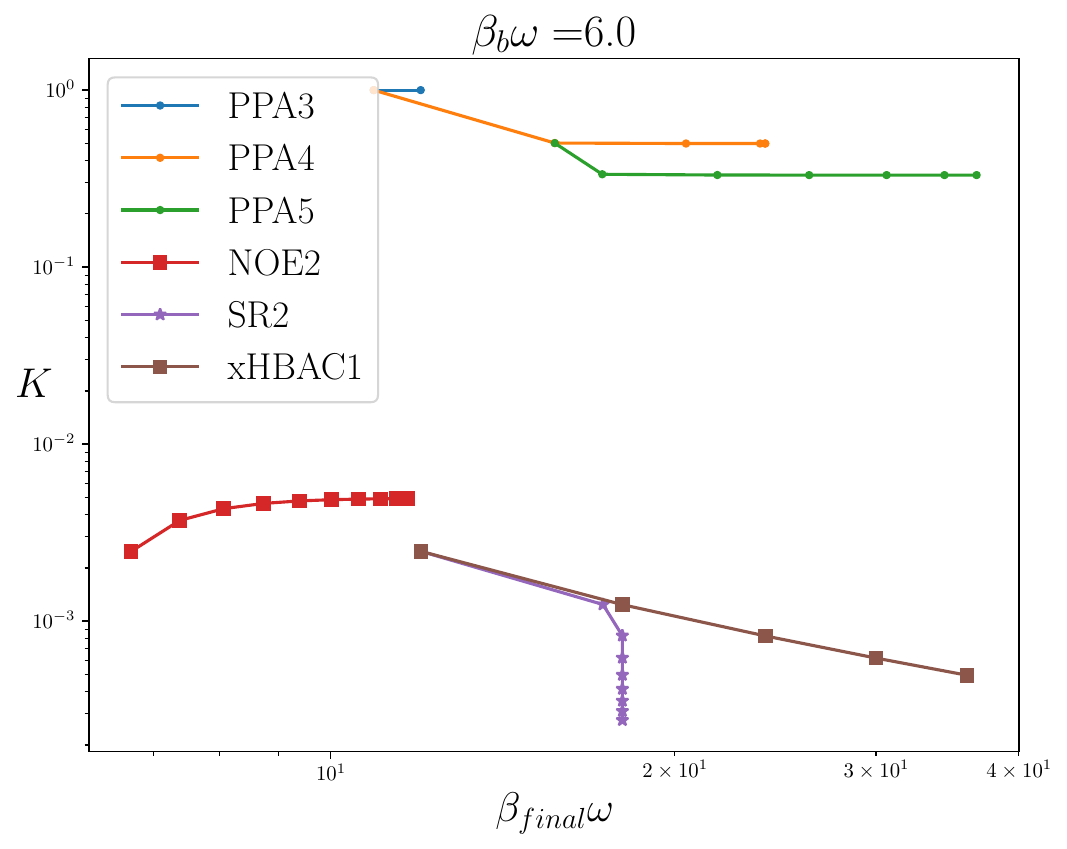}
\end{subfigure}

\caption{A plot of the cumulative COP for selected protocols starting from dimensionless inverse bath temperature $\beta_{b} \omega = 6.0$, corresponding to a situation where a qubit with energy gap $\omega/2\pi = 5 \text{GHz}$ is placed in a bath with temperature $T=40\text{mK}$.}
\label{fig_COP_comp_superconducting}
\end{figure}

As a practical example, let us consider a superconducting qubit with energy gap $\omega/2\pi = 5 \text{GHz}$ which is placed in a dilution refrigerator, initially at a bath temperature of 40mK.
This corresponds to a dimensionless temperature of
\begin{equation}
    \beta_{b} \omega \times \frac{h}{k_{B}} \approx 6.0
\end{equation}
when converted to SI units.
In \cref{fig_COP_comp_superconducting}, we plot and compare some selected protocols under this condition.
Assume now that we would like to cool the target qubit to a final temperature of $10\text{mK}$, corresponding to $\beta_{\text{final}} \omega = 24$.
In order to choose the best protocol, note that 
\begin{enumerate}
    \item The PPA3, NOE2, and SR2 have equilibrium temperatures that are higher than the target temperature
    \item The xHBAC, PPA4 and above can achieve the target temperature, among which the PPA4 does so with the highest COP, and reaches equilibrium in a small number of rounds
\end{enumerate}
Therefore, the PPA4 would be the optimal choice for this purpose, based solely on energy considerations.
For a different combination of initial and final target temperature, one can follow a similar procedure to determine the best protocol.
Typically, it is sufficient to numerically compute the COP for the PPA with the smallest number of qubits that can cool to the target temperature, and compare with the COP for the other protocols which can be calculated analytically.

\section{Heat production and the Landauer Ratio}\label{sec_Landauer}
In addition to the work required, another equally interesting aspect is to look at the heat released during a cooling procedure.
This is particularly relevant for certain quantum processing platforms that are sensitive to the environmental temperature, such as superconducting qubits.
In these platforms, one must minimize heating up of the environment while performing information processing.
Therefore, protocols that produces less heat will be more desirable to be used in practice, motivating our study on heat production.

\subsection{Minimal heat production}
Landauer~\cite{Landauer1961a} was the first to realize that classical information erasure (more precisely, logically irreversible operations) must come with heat released to the environment.
This phenomenon has been referred to as Landauer's Principle (LP) in later studies.
The original arguments can be summarized as follows.
Consider a single spin which acts as a memory.
The initial state of the spin is taken to be unknown; hence statistically it is equally probable to be in either spin up or spin down.
Invoking the Boltzmann definition of statistical entropy, $S = k \log(W)$ where $W$ is the number of possible microstates a system can be in, one sees that if one is to perform a \emph{reset} operation on the spins, then $W$ is reduced to half of its initial value.
Therefore, the entropy of the memory has decreased by $k \log(2)$.
Since the entropy of a closed system (memory plus environment) cannot decrease, one can then infer that the entropy of the environment must have increased by at least $k \log(2)$.

It is expected that a similar argument should hold when the system of interest is of quantum nature.
The analogy goes that unitary evolutions are reversible and does not change the entropy, therefore is analogous to reversible operations in classical thermodynamics.
Meanwhile, a non-unitary transformation may result in a finite entropy change, and a negative entropy change is analogous to classical information removal.
Therefore, the goal of formulating a quantum Landauer's Principle has largely been focusing on bounds of heat production in terms of entropy change in the target system.
In particular, note that the goal of algorithmic cooling is to drive the target system towards a fixed state (the ground state in our examined cases), and necessarily accompanies an entropy change.
Below we will review two inequalities of this type, then propose an efficiency measure for algorithmic cooling based on entropy considerations.

In the first scenario, there exists a target system of interest initially in an arbitrary state $\rho_{t}$, and a bath initially in a thermal state $\rho_{b}^{th}$ with inverse temperature $\beta$, jointly in a tensor product state $\rho_{t,b} = \rho_{t} \otimes \rho_{b}$.
The full state of system plus bath is thermodynamically closed.
The system and bath then undergo a joint unitary transformation denoted by $U_{t,b}$, taking them to a final state $\rho_{t,b}'$.
The work by Reeb and Wolf~\cite{Reeb2014a} showed that (see \cref{sec_appen_LP_unitary} for a derivation)
\begin{equation}\label{eqn_LP_driven}
    \beta \Delta E_{b} = - \Delta S_{t} + I(t':b') + D(\rho_{b}' || \rho_{b})
\end{equation}
where $\Delta E_{b} = \Tr[(\rho_{b}'-\rho_{b})H_{b}]$ is the energy increase of the bath, $-\Delta S_{t} = - (S(\rho_{t}') -S(\rho_{t}))$ is the entropy decrease of the system, $I(t':b')$ is the quantum mutual information between target and bath after the evolution, and $D(\rho_{b}' || \rho_{b})$ is the quantum relative entropy between initial and final states of the bath,
\begin{equation}
    \begin{gathered}
        I(t':b') = S(\Tr_{b}[\rho_{t,b}']) + S(\Tr_{t}[\rho_{t,b}']) - S(\rho_{t,b}') \\
        D(\rho_{b}' || \rho_{b}) = -S(\Tr_{t}[\rho_{t,b}']) -\Tr[\Tr_{t}[\rho_{t,b}'] \log \rho_{b}]
    \end{gathered}
\end{equation}
Due to the non-negativity of $I$ and $D$, this result implies that the energy increase in the bath (after scaled by the inverse temperature) will be lower bounded by the entropy decrease in the system.

In the second scenario, consider a single target system undergoing a general CPTP evolution $\mc{E}_{t}$.
When the system is open, $\mc{E}_{t}$ may be a non-unitary evolution.
In general we may extend this CPTP map to be a unitary map on a larger Hilbert space, but since the extension is non-unique, the exact evolution on the ancillary system is also non-unique.
Interestingly, a Landauer-type inequality can be derived for a subset of such processes, by only examining the change in the system alone.
Fro a general CPTP evolution on the target system where $\rho_{t}' = \mc{E}_{t}(\rho_{t})$, one can derive (see \cref{sec_appen_LP_non_unitary}) the following equality:
\begin{equation}\label{eqn_LP_non_unitary}
    -\beta \Delta E_{t} = -\Delta S_{t} + (D(\rho_{t} || \rho_{t}^{th}) - D(\rho_{t}' || \rho_{t}^{th}))
\end{equation}
where, as before, $-\beta \Delta E_{t} = -\Tr[(\rho_{t}'-\rho_{t})H_{t}]$ is the negative energy increase in the system, $-\Delta S_{t} = S(\rho_{t}) -S(\rho_{t}')$ is the entropy decrease in the system, and $\rho_{t}^{th}$ is the system's thermal state.

Interpreting quantum relative entropy as a distance measure, one sees that for processes where the final state is closer to $\rho_{t}^{th}$ than the initial state, the difference term in the bracket on the right is positive.
For example, this will hold true for a thermalization process towards $\rho_{t}^{th}$.
Therefore, for processes where the state is becoming closer to the thermal state, \cref{eqn_QLP_spont} states that the decrease in energy (scaled by the inverse bath temperature) will be larger than the decrease in entropy for the system.
We therefore refer to \cref{eqn_LP_non_unitary} as the LP for thermalization processes.

To relate to energy change in the bath, a common assumption (e.g., in collision models~\cite{Lorenzo2015a,Strasberg2017}) is that the actual joint evolution of the system and bath is energy-preserving.
This implies that $\beta \Delta E_{b} = - \beta \Delta E_{t}$, so that the energy \textit{increase} in the bath can now be related to the entropy decrease in the system as
\begin{equation}\label{eqn_QLP_spont}
    \beta \Delta E_{b} = -\Delta S_{t} + (D(\rho_{t} || \rho_{t}^{th}) - D(\rho_{t}' || \rho_{t}^{th})).
\end{equation}

We have seen that for both scenarios considered above, one can establish Landauer-type inequalities which lower bounds the amount of heat increase in the bath by the amount of entropy decrease in the system.
Since the goal of HBAC is exactly reducing entropy of the system, it is well motivated to define an ``entropic efficiency'' for a given protocol by considering the energy increase \textit{outside} the target system.
A slight complication for the coherent control cooling schemes is that outside the target, there exists both a machine and a heat bath where the energy can be directed to.
While the bath is assumed to be large such that it remains in a thermal state at fixed temperature, the machine may not be thermal after the first round of cooling, preventing a direct application of \cref{eqn_LP_driven} for the ``control'' step.
Nonetheless, the following quantity
\begin{equation}\label{eqn_defn_LR}
    R_{L} \coloneqq \frac{\beta_{b} (\Delta E_{b} + \Delta E_{m})}{-\Delta S_{t}}
\end{equation}
where $\beta_{b}$ is the inverse temperature of the bath, is well-defined for any nontrivial cooling procedure with $\Delta S_{t}>0$.
We call this the \textit{Landauer Ratio} (LR) and propose it as a measure of entropic efficiency, by observing that for the two settings discussed previously, this ratio is lower bounded by 1, and a lower value implies more entropy extracted from the target system per amount of energy released.
Like the CoP, we can study the LR for a particular round where the difference ($\Delta$) quantities above refer to the change for a round, or the cumulative LR where the difference quantities refer to the cumulative change.
We will denote the prior by $r_{L}$, and reserve $R_{L}$ as the cumulative LR, noting that the two quantities coincide for a single round of cooling.

\begin{figure}[ht]
\centering
\begin{subfigure}{0.96\linewidth}
    \includegraphics[width=\linewidth]{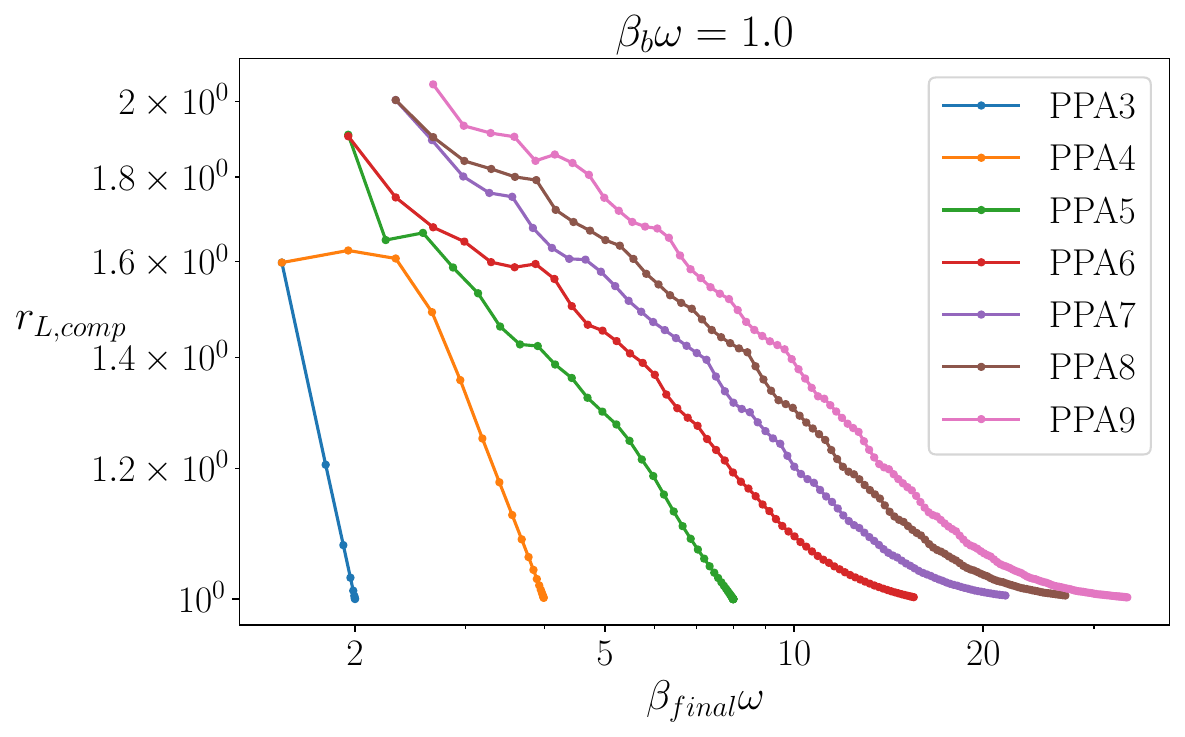}
\end{subfigure}

\caption{A plot of per-step LR, showing the ratio of the energy increase in the last two reset qubits over the entropy reduction in the target plus computational qubits, for PPA protocols. This illustrates the asymptotic optimality of PPA for cooling the computational qubits.}
\label{fig_LR_comp_PPA}
\end{figure}

\cref{eqn_defn_LR} is again an operational definition, like the CoP, because it is a ratio of cost over desired effect.
What we have in mind is the coherent cooling scheme with some predetermined target system, machine, and heat bath.
Interestingly, if we change the definition of what a target is, the LR may reveal further insights about the evolution of a cooling protocol.
A particularly insightful example is when we focus on a single unitary process, e.g., the compression step in PPA.
Specifically, we know that the purpose of compression is to concentrate entropy on the two relaxation qubits, which release those entropy to the bath during the thermalization step.
How efficient is this concentrating operation?
We first note the following theorem which is proved in \cref{sec_appen_proof_lower_LR}.

\begin{theorem}
    Consider a finite dimensional system with initial state $\rho_{xy}=\rho_{x} \otimes \rho_{y}^{th}$ where $y$ has an inverse temperature $\beta$. Let $T(\rho,\sigma)\coloneqq \frac{1}{2}\norm{\rho-\sigma}_{1}$ denote the trace distance between $\rho$ and $\sigma$. Then for a unitary process $\mc{U}(\rho_{xy})= \rho_{xy}'$ where $T(\rho_{x}',\rho_{x}) \leq 1/2e$ and $-\Delta S_{x}>0$, \begin{equation}\label{eqn_thm1}
        \frac{\beta \Delta E_{y}}{-\Delta S_{x}} \geq 1 + \frac{\gamma_{y}^{2} 
 + \lambda_{xy}^{2}}{\epsilon_{x} \log(\frac{d_{x}}{2\epsilon_{x}})}
    \end{equation} where $\epsilon_{x} = T(\rho_{x}',\rho_{x})$, $\gamma_{y} = T(\rho_{y}',\rho_{y})$, $\lambda_{xy} = T(\rho_{xy}',\rho_{x}' \otimes \rho_{y}')$, and $d_{x}$ is the dimension of subsystem $x$.
\end{theorem}

Clearly, this is exactly the setting described prior to \cref{eqn_LP_driven}, where a bipartite unitary process occurs from an initially tensor product state, with the second subsystem in the thermal state.
Therefore we expect the ratio on the left to be greater than or equal to 1.
Meanwhile, this theorem provides some insights into when the left-hand side (LHS) may approach the theoretical minimum of 1.
Specifically, this has to do with how ``drastic'' the unitary is, how ``evenly distributed'' the change is between subsystems $x$ and $y$, and how much mutual information has been created.
We observe that the lower bound would approach 1 for a unitary that satisfy the following:
\begin{enumerate}
    \item It results in a marginally small variation in the joint state $\rho_{xy}$ (when measured by the trace distance $T$), so that all of $\epsilon_{x}$, $\gamma_{y}$, and $\lambda_{xy}$ are small;
    \item The change in subsystem $x$, $\epsilon_{x}$, is less than or on the same order as the change in subsystem $y$, $\gamma_{y}$, and $\lambda_{xy}$ which is proportional to the square root of the mutual information created by the unitary.
\end{enumerate}
One could see that if the above are true, then the numerator in the extra term on the right-hand side (RHS) goes quadratic in the small quantities, while the denominator only goes as $\epsilon \log(\epsilon)$, so the ratio would go towards 0.
Of course, this is only a rough intuition, and we shall now see how this lower bound is asymptotically achieved in the PPA protocol.
In addition, we note here that a tighter but more sophisticated bound has been derived in \cite{Reeb2014a}.
The significance of our result is to give an intuition for when the lower bound of $1$ may be saturated, in terms of the change and correlation quantities as a result of the unitary.

To relate back to PPA, notice that immediately before the compression step of each round, the first $n-2$ qubits are in a tensor product state with the last two qubits, which have just been refreshed to the bath temperature.
Therefore, we may take the first $n-2$ qubits to be subsystem $x$ and the last two qubits to be subsystem $y$, so that \cref{eqn_thm1} is satisfied.
Comparing with \cref{eqn_defn_LR}, one sees that the LHS of \cref{eqn_thm1} is precisely $R_{L}$ where $\Delta E_{b}=0$ (since the system is closed during the unitary), if we identify the target system to be the first $n-2$ qubits, and the machine to be the last 2 qubits.
We will refer to the LR with this identification as $r_{L,\text{comp}}$ since it is a per-round LR where the new ``target'' system consists of all \textit{computational} qubits (a term used in literature~\cite{Rodriguez-Briones2016} to refer to all but the relaxation qubits).
What this represents is the ability of PPA to extract entropy from the computational qubits, and concentrate them into the two relaxation qubits.
A ratio close to 1 implies maximum entropy extracted from the computational qubits, per energy increase in the relaxation qubits, by the compression unitary.
We plot this quantity $r_{L,\text{comp}}$ in \cref{fig_LR_comp_PPA}, for different PPA protocols.
As can be seen from the plot, all the protocols indeed show a consistent behavior where the $r_{L,\text{comp}}$ goes towards 1 as cooling proceeds towards equilibrium.
Therefore, the PPA compression subroutine is asymptotically optimal in terms of extracting entropy from the computational qubits.

\subsection{Relating to the coefficient of performance}
Having operationally defined the Landauer Ratio, we then ask the same question of whether it can be computed given a protocol.
The difficulty lies in computing $\Delta E_{b}$ for a general process, since the bath degree of freedom is typically not  known in full detail, whereas the protocol only dictates how the system (target and machine) evolves.
Meanwhile, we will see that for coherent cooling protocols, the LR can be calculated in principle.
For the control subroutines, the joint system of target plus machine is closed, and undergoes a unitary transformation.
The quantity $\Delta E_{b}$ is thus $0$ for these processes.
Meanwhile, for the thermalization subroutines, we again assume that it occurs with a joint energy-preserving unitary on the system and bath, so that $\Delta E_{b} = -(\Delta E_{t} + \Delta E_{m})$.
Therefore, for coherent control protocols,
\begin{equation}
    \Delta E_{b} = -(\Delta E_{t,\alpha} + \Delta E_{m,\alpha})
\end{equation}
where $\alpha$ denotes all thermalization steps during the protocol.

It is instructive to see how the LR relates to the CoP defined in the previous section.
Since the full system of (target, machine, bath) is closed, energy conservation implies that 
\begin{equation}
    \Delta E_{t} + \Delta E_{m} + \Delta E_{b} = W.
\end{equation}
Replacing $W$ in the definition of the CoP gives
\begin{equation}
    K = \frac{-\Delta E_{t}}{W} = \frac{-\Delta E_{t}}{\Delta E_{t} + \Delta E_{m} + \Delta E_{b}},
\end{equation}
and rearranging the terms gives
\begin{equation}
    \frac{\Delta E_{b} + \Delta E_{m}}{\Delta E_{t}} = -\left(\frac{1}{K} + 1 \right).
\end{equation}
Recalling the definition of $R_{L}$, we have
\begin{equation}\label{eqn_COP_RL}
    R_{L} = \frac{\beta_{b} (\Delta E_{b} + \Delta E_{m})}{-\Delta S_{t}} = \frac{-\beta_{b} \Delta E_{t}}{-\Delta S_{t}} \left(\frac{1}{K} + 1 \right).
\end{equation}
This above equation shows how $K$ and $R_{L}$ are related to each other.
For a given cooling procedure with known initial and final states, knowing $K$ immediately gives $R_{L}$, and vice versa.
Moreover, it can be seen that for a given process with fixed initial and final states, a higher CoP corresponds to a lower LR, confirming our previous intuition about what ``more efficient'' means under both measures.
An equivalent way of saying this is that a higher work input necessarily accompanies more heat released into the (machine and) bath.

When the target system is a qubit with energy gap $\omega$, the cumulative LR up to the $N$-th round is simplified to
\begin{equation}\label{eqn_LR_COP}
    R_{L}(N) = \frac{\beta_{b} \omega (p_{t}(N)-p_{t}(0))}{h(p_{t}(0))-h(p_{t}(N))} (\frac{1}{K(N)}+1).
\end{equation}
where $h(p) = -p \log(p) - (1-p) \log(1-p)$.
This applies to all cases studied in this work.
We summarize the results in \cref{tab_efficiencies} which can be used to directly obtain the analytical expression for $R_{L}$.
We omit the full expressions of $R_{L}$ for brevity and show plots for them below.
In addition, for completeness, the per-round Landauer Ratio $r_{L}$ is obtained by choosing the process to be a single round of cooling in \cref{eqn_defn_LR}.
For coherent cooling procedures, we can modify \cref{eqn_LR_COP} to obtain
\begin{equation}
    r_{L}(N) = \frac{\beta_{b} \omega (p_{t}(N)-p_{t}(N-1))}{h(p_{t}(N-1))-h(p_{t}(N))} (\frac{1}{k(N)}+1)
\end{equation}
as the Landauer Ratio for the $N$-th round.

\begin{figure}
\centering
\begin{subfigure}{0.96\linewidth}
    \includegraphics[width=\linewidth]{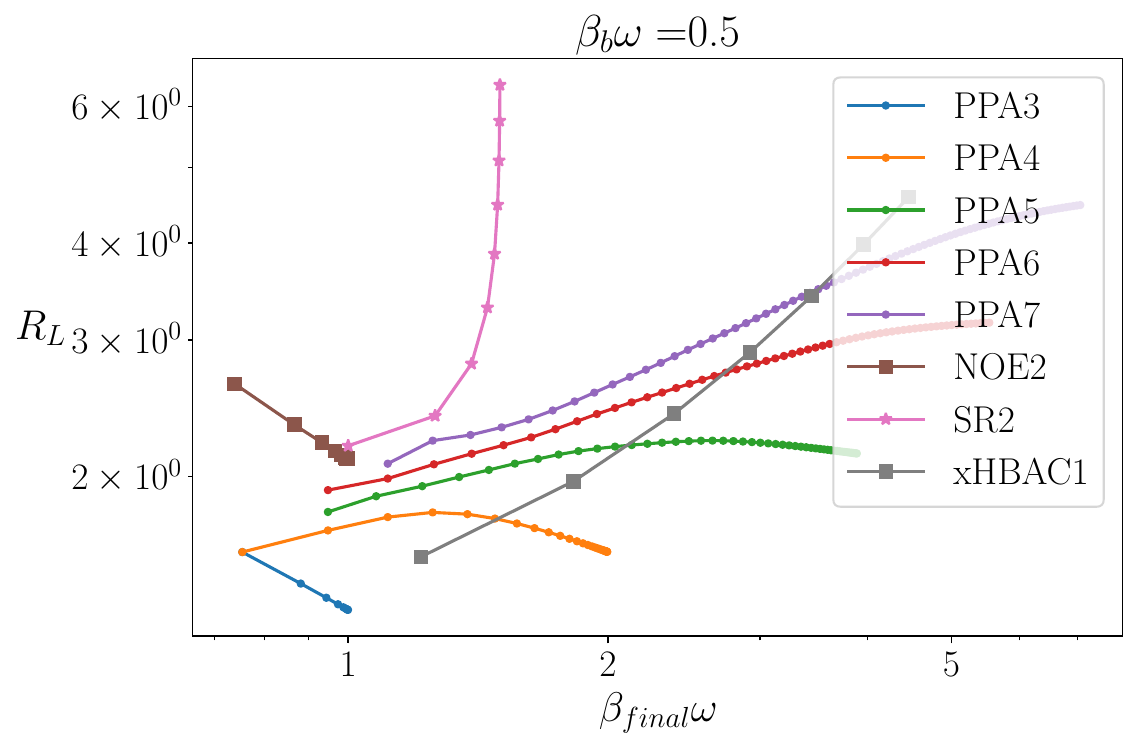}
\end{subfigure}
\hfill
\begin{subfigure}{0.96\linewidth}
    \includegraphics[width=\linewidth]{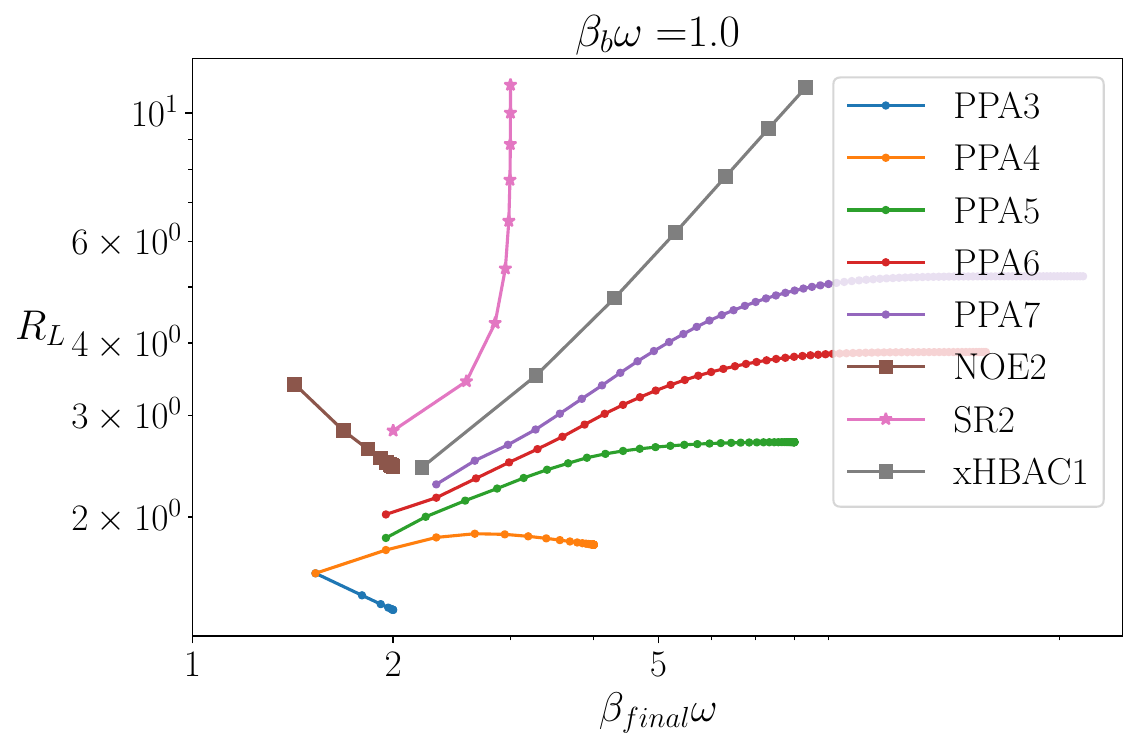}
\end{subfigure}
\hfill
\begin{subfigure}{0.96\linewidth}
    \includegraphics[width=\linewidth]{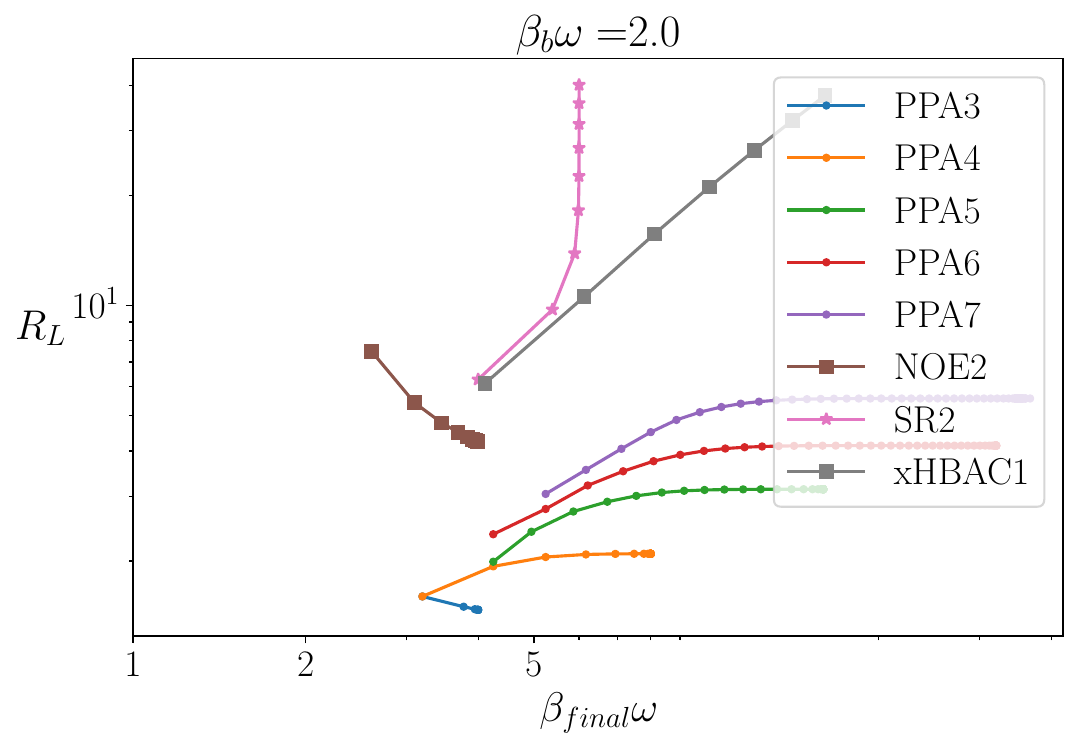}
\end{subfigure}
\caption{Comparison of cumulative LR between HBAC protocols under different initial bath temperatures, plotted against the final inverse temperature $\beta_{\text{final}} \omega$.}
\label{fig_LR}
\end{figure}

In \cref{fig_LR}, we show a plot of $R_{L}$ for different HBAC protocols.
Comparing with \cref{fig_COP} at the same bath temperature, we may verify our result in \cref{eqn_COP_RL} that protocols with a lower CoP necessarily have a higher LR, so both are meaningful figures of merit for efficiency.
At $\beta_{b} \omega = 1$ and above, the PPA protocols under consideration releases less amount of heat into the environment when cooling to the same final target temperature.
Meanwhile, at a higher temperature of $\beta_{b} \omega = 0.5$, there are certain regions in which using the xHBAC1 is more efficient than the PPAs, when both can reach the same final temperature.

Interestingly, note that both the PPA3 and NOE have LR's that decrease with further cooling.
This can be understood from \cref{eqn_COP_RL} that, for $K$ being fixed, the LR is proportional to the ratio of energy change over entropy change in the target system.
When the target is a single qubit, both changes are functions of a single parameter $p$.
Since the rate of decrease in energy (linear in $p$) is faster than the rate of decrease in entropy (log in $p$) as $p \rightarrow 1$, the ratio will grow smaller as $p$ increases.
Therefore, for protocols like the PPA3 and NOE which have constant or near-constant $K$, the change in LR is dominated by this effect and decreases with the cooling progress.

\begin{table*}[ht]
\renewcommand*{\arraystretch}{2.8}
\centering
\begin{tabular}{|c|c|c|c|}
\hline
 & $k(N)$ & $K(N)$ & $p_{t}(N)$ \\ \hline
$\ppa$ & $1$ & $1$ & $\displaystyle \frac{z}{1+z} - y^{N} (\frac{z}{1+z} - p_{t}(0))$ \\ [1ex] \hline
$\noe$ & $\displaystyle 
        \frac{p_{b2}-p_{b}}{2 p_{b}-1} \text{ if } N=1,\ 
        \frac{1}{2} \text{ if } N>1 $ & $\displaystyle \frac{(p_{b2}-p_{b}) (1-2^{-N})}{(p_{b2} - \frac{1}{2}) - 2^{1-N} (p_{b2} - p_{b})}$ & $\displaystyle \frac{x}{1+x} - 2^{N} (\frac{x}{1+x} - p_{t}(0))$ \\ [1ex] \hline
$\srg$ & \quad $\displaystyle  \frac{p_{b} (1 - p_{b}) (2p_{b2}-1)}{2p_{b}-1} (p_{b2}+p_{b}-2p_{b2}p_{b})^{N-1}$ \quad & \quad  \quad $\displaystyle \frac{p_{b} (1 - p_{b}) (2p_{b2}-1)}{2p_{b}-1} \frac{1-u^{N}}{N(1-u)}$ \quad \quad & \quad$\displaystyle \frac{w}{w+v} - u^{N} (\frac{w}{w+v} - p_{t}(0))$ \quad \\ [0.5ex] \hline
$\xhbac$ & $\displaystyle \frac{(1 - p_{b})(1-s)}{1-2s^{N-1}(1-p_{b})} s^{N-1}$ & $\displaystyle \frac{(1-p_{b})(1-s)(1-s^{N})}{n(1-s)-2(1-p_{b}) (1-s^{N})}$ & $\displaystyle 1 - s^{N} (1 - p_{t}(0))$ \\ [0.8ex] \hline
\end{tabular}
\caption{Summary of analytical expressions for the per-round and cumulative coefficients, and ground-state population evolution of performance for different protocols studied. A summary for the shorthand notations are as follows: $x=\frac{p_{b2}}{1-p_{b2}}$, $y \coloneqq  2p_{b}(1-p_{b})$, $z \coloneqq p_{b}^{2}/(1-p_{b})^{2}$, $w \coloneqq p_{b2} p_{b}$, $v = 1-p_{b2} -p_{b} + p_{b2} p_{b}$, $u = 1-v-w$, $s=e^{-\beta_{b}\omega}$.}
\label{tab_efficiencies}
\end{table*}

\section{Designing protocols with better energy efficiency}\label{sec_improved_HBAC}
We noted previously that conventionally, analyses on HBAC have mostly focused on the cooling speed and asymptotic achievable temperature.
This drives searches for faster protocols or ones which can cool to lower temperatures, but largely ignores the amount of energy cost.
Having taken energy aspects into consideration for HBAC protocols, we now study possible methods to improve their energy performance.
To be more concrete, we will first look at the compression step in PPA, and show how it can be improved energetically.
Later, we will see how this idea leads to more efficient variants of the PPA and xHBAC protocols.

Consider the same setup as the PPA which consists of a target system with $X$ energy levels and a machine with $Y$ energy levels, with the possibility of having degeneracies.
For simplicity, we assume that both the target and the ancillas are diagonal in the energy eigenbasis.
In such systems, a unitary corresponds to permutations of the diagonal elements (up to a global phase), so it is sufficient to consider the simpler problem of optimizing over all possible permutations.
We ask the question: among the unitaries that achieve maximal energy reduction in the target, which is the most energy-efficient one?
Note that here we do not try to optimize the CoP, and instead, only optimize within the set of unitaries that achieve the most cooling possible given the initial system.

\subsection{Improved PPA}
We first examine the original PPA entropy compression, which performs a descending SORT to the diagonal elements with respect to the lexicographic basis order.
For simplicity, we will assume that all qubits have the same energy gap.
Recalling that the target system in PPA is a single qubit, so $X=2$.
Therefore, the descending SORT always achieve the maximal energy reduction in the target, since it maximally populates the lowest energy ground state.
Therefore, one only needs to look at the energy cost of this operation and check whether there exists a more efficient one, whose existence we will show below.

For PPA3, the states in the $\ket{0}$ subspace of the first qubit include
\begin{equation*}
    000,\ 001,\ 010,\ 011,
\end{equation*}
and similarly for the $\ket{1}$ subspace.
We see that these states are already ordered in terms of energy assuming the same energy gap between all qubits, so a descending SORT would result in the lowest energy configuration within each subspace of the target.

In contrast, for $n = 4$ qubits, the states in the $\ket{0}$ subspace of the first qubit are
\begin{equation*}
    0000,\ 0001,\ 0010,\ 0011,\ 0100,\ 0101,\ 0110,\ 0111
\end{equation*}
and similarly for the $\ket{1}$ subspace.
These states are not fully ordered in terms of energy: specifically the state $\ket{0011}$ comes before $\ket{0100}$, but has a higher energy.
Consider the following new ordering
\begin{equation*}
    0000,\ 0001,\ 0010,\ 0100,\ 0011,\ 0101,\ 0110,\ 0111
\end{equation*}
which is now ordered in ascending energy within the target $\ket{0}$ subspace.
Compared with the original compression, a descending SORT with respect to this new basis order achieves the same cooling in the target, but uses potentially less energy since the low-energy state $0100$ is now more populated than the high-energy state $0011$.


We now go back to the more general setting, where we wish to reduce the energy in an $X$-level target system as much as possible with a single unitary.
The available ancillary system is a machine with $Y$ energy levels, such that the full system has a total of $XY$ levels.
Following the previous intuition, we propose the following compression procedure:

\begin{enumerate}
    \item Label the target subspaces in non-decreasing order, such that $E_{t,1} \leq \dots \leq E_{t,X}$;
    \item Label the machine subspaces in non-decreasing order, such that $E_{m,1} \leq \dots \leq E_{m,Y}$. This results in an ordering for the full system as 
    \begin{equation*}
        \ket{1,1},\dots,\ket{1,Y}, \ket{2,1},\dots, \ket{2,Y},\dots,\ket{X,Y}
    \end{equation*}
    where the first index labels the target and the second labels the machine;
    \item Perform a descending SORT with respect to this basis ordering, so that the population profile after the unitary satisfies:
    \begin{equation*}
        p_{1,1} \geq \dots \geq p_{1,Y} \geq p_{2,1} \geq \dots \geq p_{X,Y}.
    \end{equation*}
\end{enumerate}

\begin{figure}
\centering
\begin{subfigure}{0.96\linewidth}
    \includegraphics[width=\linewidth]{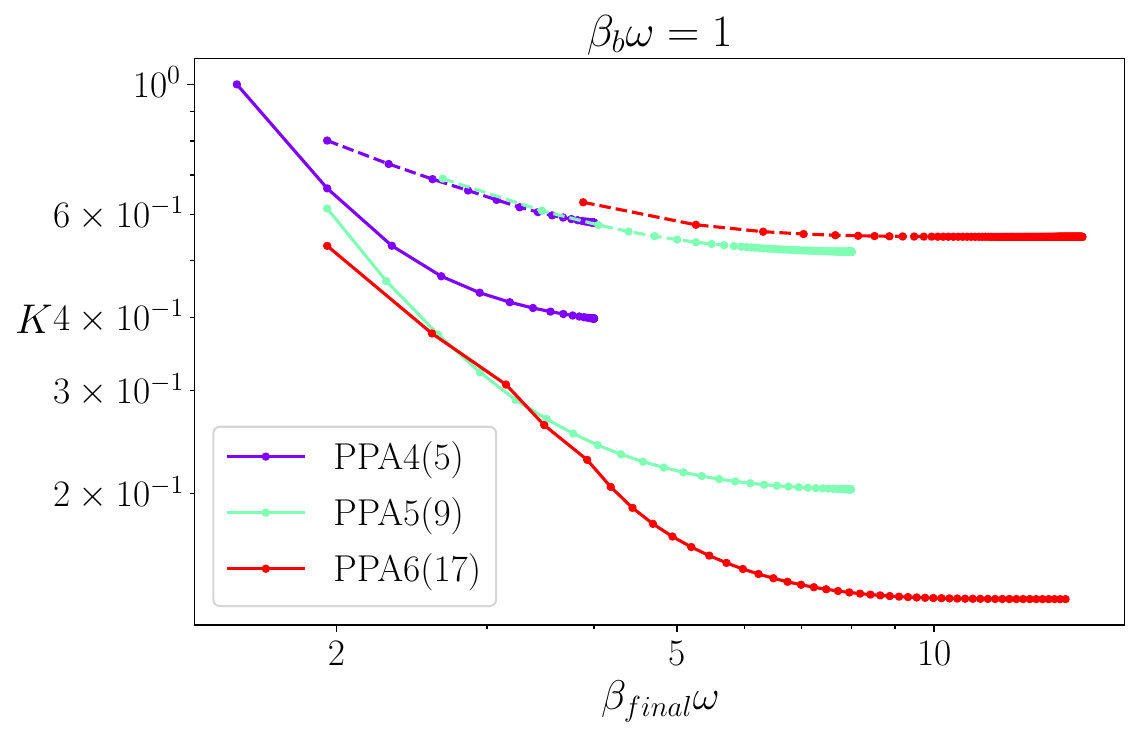}
\end{subfigure}

\caption{Comparison of CoP between the original (solid lines) and energetically improved (dashed lines) PPA protocols. The improved PPA uses a larger number of qubits (indicated in the brackets in the plot legend) compared with the original PPA in order to cool the target to the same asymptotic temperature, but uses less energy in total to achieve so.}
\label{fig_PPA_improved}
\end{figure}

The above procedure ensures maximal energy reduction in the target, while requiring the least amount of work input, in a single unitary step.
It is interesting to see how this affects the performance of PPA, when combined with non-unitary relaxation steps in the original protocol.
A round in our new protocol consists of: 1) a control subroutine using the above compression procedure, and 2) a thermalization subroutine where all qubits higher than the bath temperature are relaxed to the bath temperature $T_{b}$.
Interestingly, starting from an all-thermal state at $T_{b}$, we find that all qubits except the target becomes warmer than the bath, and are therefore all relaxed after the new compression.
According to \cite{Rodriguez-Briones2016}, relaxing more qubits implies a higher achievable temperature of the target.
In particular, we show in \cref{appen_proof_efficient_PPA} that the asymptotic achievable temperature using this new procedure is $T_{b}/(n-1)$ with a total of $n$ qubits, compared to $T_{b}/2^{n-2}$ in the original PPA.
Consequently, a fair CoP comparison is between the new scheme using $2^{n-2}+1$ qubits and the original one using $n$ qubits, such that both achieve the same asymptotic temperature.
This is shown in \cref{fig_PPA_improved} where we compared the first three instances, and see indeed that the improved protocol has a higher cumulative CoP than the original one.

We end with a brief discussion on the above results.
Compared with the original scheme, the new scheme creates a more ``averaged'' compression, such that the temperatures of qubits other than the target are more evenly distributed.
This ``softer'' compression uses less energy, but also creates less temperature gradient among the machine bits.
Thus, this change the structure of the asymptotic state of the full system: while that of the original PPA has the following temperature profile
\begin{equation*}
    \frac{T_{b}}{2^{n-2}},\ \frac{T_{b}}{2^{n-3}},\ \dots,\ \frac{T_{b}}{2},\ T_{b},\ T_{b}
\end{equation*}
for each qubit, the new PPA has
\begin{equation*}
    \frac{T_{b}}{n-1},\ T_{b},\ T_{b},\ \dots, T_{b}.
\end{equation*}

One can see this as a way to trade spatial cost with energetic cost: in order to create the temperature gradient in the original PPA, it is necessary to cool the first $n-2$ qubits to temperatures lower than the bath.
The new PPA avoids such unnecessary cooling, by using a larger total number of qubits but with much lower energy cost.

\subsection{Improved xHBAC}
Next, we see how the same principle can be used to obtain an improved realization of the xHBAC1 protocol, compared with the original proposal which uses a bosonic bath interacting with the qubit to implement the $\Lambda_{\beta}$ operation~\cite{Alhambra2019}.
While the original proposal can already cool down the target qubit exponentially fast, as can be seen from \cref{fig_COP}, it is not thermodynamically optimal in terms of energy consumption, since the total amount of work input is unbounded as more cooling rounds are performed.
In fact, it is well known that using an ancillary system with an unbounded spectrum, a target system can be cooled to zero temperature using unitary control.
Using this same ancillary system, we can analytically construct the final state of the combined system and calculate the work cost.

Again, the target qubit and the harmonic oscillator start in a tensor product state, each being in a thermal state at the same inverse temperature $\beta_{b}$.
This corresponds to a situation with $X=2$ and $Y=\infty$.
We will consider the ``resonant'' scenario where $E_{t} = E_{a} \coloneqq \omega$, since this is the necessary condition for $\Lambda_{\beta}$ to be a thermal operation~\cite{Alhambra2019}.
Then, the thermal population of a state $\ket{m, n}$ only depends on the ``total quantum number'' $m+n$: for example, the population of $\ket{1,0}$ and that of $\ket{0,1}$ will be identical.
Moreover, the ordering for the full system will be given by
\begin{equation*}
    \ket{0,0},\ \ket{0,1},\ \ket{0,2},\ \ket{0,3},\ \dots
\end{equation*}
where the first index labels the qubit and the second labels the bosonic mode.
Thus, the improved protocol puts the largest diagonal elements in the joint density matrix into the $\ket{0}$ subspace of the qubit, in decreasing order of magnitude with the increasing energy levels of the harmonic oscillator $\ket{n}$.
Due to the infinite energy levels present in the harmonic oscillator, the final state would have all the significant components in the qubit $\ket{0}$ subspace, which can be formally seen by first considering a truncated $Y$-level machine, and taking the limit $Y \rightarrow \infty$.

Therefore, after the optimized compression, the final state would have the $\ket{0,0}$ population unchanged, then the original $\ket{0,i}$ population ``copied twice'' into $\ket{0,2i-1}$ and $\ket{0,2i}$, for $i=1 \dots \infty$.
The energy of this final state is given by 
\begin{equation}
\begin{aligned}
    E_{\text{fin}} &= \sum_{i=1}^{\infty} \frac{1}{Z_{a}} p_{0} e^{-i \beta_{b} \omega} (2i+2i-1)\omega\\
    =& \sum_{i=1}^{\infty} (1-e^{-\beta_{b} \omega}) (\frac{1}{1+e^{-\beta_{b} \omega}}) e^{-i \beta_{b} \omega} (4i-1)\omega\\
    =& \frac{1+3e^{\beta_{b} \omega}}{e^{2\beta_{b} \omega}-1} \omega
\end{aligned}
\end{equation}
where the ground state $\ket{0,0}$ has been set to have energy $0$.
The energy of the initial state is simply the energy of the qubit plus that of the harmonic oscillator,
\begin{equation}
    E_{\text{ini}} = \delta_{b} \omega + \sum_{i=0}^{\infty} \frac{1}{Z_{a}} e^{-i \beta_{b} \omega} i \omega = \frac{2e^{\beta_{b} \omega}}{e^{2\beta_{b} \omega}-1} \omega
\end{equation}
The work input is
\begin{equation}
    W = E_{\text{fin}} - E_{\text{ini}} = \frac{1}{e^{\beta_{b} \omega}-1} \omega.
\end{equation}
Since the final energy of the qubit is 0, the energy decrease in the qubit is the initial energy of the qubit, $\delta_{b} \omega$.
The CoP is thus
\begin{equation}
    K = \frac{-\Delta E_{t}}{W} = \frac{1-e^{-\beta_{b} \omega}}{1+e^{-\beta_{b} \omega}} = \tanh{\frac{\beta_{b} \omega}{2}}.
\end{equation}

We see that the optimal cumulative CoP is bounded between $0$ and $1$, depending on the bath temperature.
Compared with the exponentially decreasing CoP (see \cref{fig_COP}) using the original proposal, this result suggests that there is much room for improvement in the energy efficiency when designing practical cooling protocols using a bosonic bath.

\section{Conclusions}

In this paper, we investigated the thermodynamic properties of algorithmic cooling within a comprehensive framework of coherent cooling, encompassing a range of distinct protocols. 
Employing the transition matrix formalism, we identified a consistent behavior among these protocols in terms of their cooling limits and target state evolution. 
Meanwhile, their thermodynamic performance displayed markedly diverse characteristics.

To compare the protocols, we proposed two key metrics: the coefficient of performance and the Landauer ratio, each highlighting a different aspect of a protocol. 
We benchmarked each protocol using both metrics, and provided guidelines for selecting the energetically optimal protocol given an initial and final temperature combination.
As a practical example, we examined the case of a superconducting qubit at milli-Kelvin temperature, and selected the 4-qubit PPA as the optimal protocol from an energy consideration standpoint.
Furthermore, we demonstrated that the PPA becomes asymptotically efficient in terms of the Landauer ratio. 
Finally, we introduced more energetically efficient cooling algorithms, inspired by both the PPA and xHBAC protocols and our thermodynamic analysis.
These improvements allow for higher energy efficiencies  by cooling to the same final temperature on the target qubit using less energy input, as compared with the original proposals.
Our findings contribute to the analysis and enhancement of the thermodynamic aspects of algorithmic cooling, as well as other quantum thermodynamic protocols.

\acknowledgments 
J.L. and N.A.R.-B contributed equally to this work.
R.L. thanks Mike and Ophelia Lazaridis for funding. N.A.R.-B. acknowledges the support from the Miller Institute for Basic Research in Science, at the University of California, Berkeley. 

\section{Appendices}
\appendix
\section{Derivation for Landauer's Principle in the unitary scenario}\label{sec_appen_LP_unitary}
This is a more concise version of the proof in \cite{Reeb2014a}.
We first prove the case where $\beta$ is finite.
Due to the invariance of the von Neumann entropy under unitary transformations, $S(\rho_{t,b}')=S(\rho_{t,b})=S(\rho_{t})+S(\rho_{b})$.
Expand the mutual information term as
\begin{equation}
    \begin{aligned}
        I(t':b') &= S(\rho_{t}')+S(\rho_{b}')-S(\rho_{t,b}')\\
        &=S(\rho_{t}')+S(\rho_{b}')-S(\rho_{t,b})\\
        &=S(\rho_{t}')+S(\rho_{b}')-S(\rho_{t})-S(\rho_{b})\\
        &= \Delta S_{t}+S(\rho_{b}')-S(\rho_{b}).
    \end{aligned}
\end{equation}
Also
\begin{equation}
\begin{aligned}
    D(\rho_{b}'|| \rho_{b}) &= \Tr[\rho_{b}' \log \rho_{b}'] - \Tr[\rho_{b}' \log \rho_{b}]\\
    &= -S(\rho_{b}')- \Tr[\rho_{b}' \log \rho_{b}]
\end{aligned}
\end{equation}
Therefore
\begin{equation}
    \begin{aligned}
        -\Delta S_{t}+I(t':b') + D(\rho_{b}'|| \rho_{b}) &= -S(\rho_{b})- \Tr[\rho_{b}' \log \rho_{b}]\\
        &= \Tr[(\rho_{b}-\rho_{b}')\log \rho_{b}].
    \end{aligned}
\end{equation}
The bath is initially in a thermal state, $\rho_{b} = e^{-\beta_{b} H_{b}}/\Tr[e^{-\beta_{b} H_{b}}]$, so
\begin{equation}
    \log \rho_{b} = -(\beta_{b} H_{b} + \mathds{1} \log \Tr[e^{-\beta_{b} H_{b}}]).
\end{equation}
Plug this back into one equation above gives
\begin{equation}\label{eqn_appen_LP_unitary}
    -\Delta S_{t}+I(t':b') + D(\rho_{b}'|| \rho_{b}) = \beta_{b}\Tr[(\rho_{b}-\rho_{b}')H_{b}]
\end{equation}
since $\rho_{b}-\rho_{b}'$ is traceless.
We can identify this to be the energy change in the bath, $\Delta E_{b}$.

For the case where $\beta = \infty$, $\rho_{b} = P_{g}/\dim(P_{g})$ is the normalized projector onto the ground state space of $H_{b}$.
Thus $\Delta E_{b} \geq 0$ and there are two possibilities:
\begin{enumerate}
    \item $\Delta E_{b}=0$, then $\rho_{b}'$ is also supported only in the ground state space of $H_{b}$.
    The key is to realize that $\log \rho_{b} = \log P_{g} - \log \dim(P_{g})$, and that $\log P_{g}$ is the $0$ operator on the support of $P_{g}$.
    Therefore
    \begin{equation}
        \begin{aligned}
            -\Delta S_{t} &+I(t':b') + D(\rho_{b}'|| \rho_{b})  = -S(\rho_{b}) - \Tr[\rho_{b}' \log \rho_{b}]\\
            &= \Tr[(\rho_{b} - \rho_{b}') \log \rho_{b}]\\
            &= \Tr[(\rho_{b} - \rho_{b}') \log P_{g}] - \Tr[(\rho_{b} - \rho_{b}')\log \dim(P_{g}))]\\
            &= 0
        \end{aligned}
    \end{equation}
    where the first term is 0 because both $\rho_{b}$ and$\rho_{b}'$ are supported only on the ground state space of $H_{b}$, and the second term is 0 because $\rho_{b} - \rho_{b}'$ is traceless.
    Thus, both sides of \cref{eqn_appen_LP_unitary} are 0.
    
    \item $\Delta E_{b}>0$, then $\rho_{b}'$ is supported outside the ground state of $H_{b}$.
    Thus $D(\rho_{b}'|| \rho_{b}) = \infty$, and both sides of \cref{eqn_appen_LP_unitary} are $\infty$. 
\end{enumerate}

\section{Derivation for Landauer's Principle in the CPTP scenario}\label{sec_appen_LP_non_unitary}
The derivation can be most easily seen by expressing the difference in relative entropy using its definition,
\begin{equation}
\begin{aligned}
    &D(\rho_{t} || \rho_{t}^{th}) - D(\rho_{t}' || \rho_{t}^{th})\\ &= - S(\rho_{t}') - \Tr[\rho_{t}' \log(\rho_{t}^{th})] + S(\rho_{t}) + \Tr[\rho_{t} \log(\rho_{t}^{th})]\\
    &= (S(\rho_{t})-S(\rho_{t}')) + \Tr[(\rho_{t} - \rho_{t}') \log(\rho_{t}^{th})]\\
    &= -\Delta S_{t} + \Tr[(\rho_{t} - \rho_{t}') \log(\rho_{t}^{th})].
\end{aligned}
\end{equation}
By definition, $\rho_{t}^{th} = e^{-\beta H_{t}}/\Tr[e^{-\beta H_{t}}]$, then
\begin{equation*}
\begin{aligned}
  &D(\rho_{t} || \rho_{t}^{th}) - D(\rho_{t}' || \rho_{t}^{th})\\
     &= -\Delta S_{t} + \Tr[(\rho_{t} - \rho_{t}') (-\beta H_{t} - \mathds{1} \log \Tr[e^{-\beta H_{t}}])]\\
    &= -\Delta S_{t} - \Tr[(\rho_{t} - \rho_{t}')\beta H_{t}] - \log \Tr[e^{-\beta H_{t}}] \Tr[(\rho_{t} - \rho_{t}')]
\end{aligned}
\end{equation*}
The last term is 0 since all physical processes are trace-preserving.
By defining $\Delta E_{t} \coloneqq \Tr[(\rho_{t}' - \rho_{t}) H_{t}]$ and $\Delta S_{t} \coloneqq S(\rho_{t}')-S(\rho_{t})$, we obtain that
\begin{equation}
    -\beta \Delta E_{t} = -\Delta S_{t} + (D(\rho_{t} || \rho_{t}^{th}) - D(\rho_{t}' || \rho_{t}^{th})).
\end{equation}

\section{Proof for lower bound on LR in unitary processes}\label{sec_appen_proof_lower_LR}
Consider the unitary evolution $\mc{U}(\rho_{ab}) = \rho_{ab}'$ which results in a negative entropy change on subsystem $a$, $-\Delta S_{a}>0$.
We can rearrange \cref{eqn_LP_driven} and divide by $-\Delta S_{a}$ to obtain an expression for the Landauer Ratio for a round of unitary operation:
\begin{equation}\label{eqn_appen_rL}
    R_{L} = 1 + \frac{I(a':b')+D(\rho_{b}' || \rho_{b})}{-\Delta S_{a}}.
\end{equation}
First note that the mutual information term $I(a':b')$ can be rewritten as (see, e.g., Eq. (3.55) in \cite{Witten2020})
\begin{equation}
    I(a':b') = D(\rho_{ab}' || \rho_{a}' \otimes \rho_{b}')
\end{equation}
A quadratic lower bound on the relative entropy is~\cite{Audenaert2005}
\begin{equation}
    D(\rho || \sigma) \geq 2(T(\rho,\sigma))^{2}
\end{equation}
On the other hand, an upper bound on the entropy change is given by the Fannes' inequality~\cite{Fannes1973,Nielsen2012}:
\begin{equation}
    \abs{S(\rho)-S(\sigma)} \leq 2 T(\rho,\sigma) \log(\frac{d}{2T(\rho,\sigma)})
\end{equation}
for $T(\rho_{a}',\rho_{a}) \leq 1/2e$.
Directly plugging in the above into \cref{eqn_appen_rL} gives the desired bound on $R_{L}$.

\section{Landauer's principle, bound for the PPA3-HBAC protocol}



At inverse temperature $\beta_{b}$, the Laundauer's principle tells us that the change of entropy in the target qubit will cause heat dissipation to the bath,
\begin{align}
    Q\geq -\Delta S_{t}/\beta_{b}
\end{align}
where $-\Delta S_{t}$ denotes the entropy decrease in the target.

For our case, starting with the PPA3, the change of entropy in the target qubit after N rounds will be

\begin{align}
    -\Delta S_{t} &=S(\epsilon_{b})-S(\epsilon_{N})\\
    &=\frac{1}{2}\sum_{j=1}^{\infty}\frac{\epsilon_{N}^{2j}-\epsilon_{b}^{2j}}{j\left(2j-1\right)}
\end{align}

And the heat released to the bath during the $N$ rounds, given by 
\begin{equation}
    Q_{\text{tot}}(N)=\omega\left(1-r^N\right)\left(\epsilon_\infty-\epsilon_{b}\right)
\end{equation}
where $\omega$ is the energy gap of all 3 qubits, and $\epsilon_{\infty}$ is calculated in \cref{sec_unified_evolution}.
Therefore, for $N$ rounds:
\begin{align}
\frac{\beta_{b} Q_{\text{tot}}(N)}{-\Delta S_{t}}=\frac{4\beta_{b} \omega\left(1-r^N\right)\left(\epsilon_\infty-\epsilon_{b}\right)}{\sum_{j=1}^{\infty}\frac{\epsilon_{N}^{2j}-\epsilon_{b}^{2j}}{j\left(2j-1\right)}}
\end{align}
where $\beta_{b}=2\omega^{-1}{\rm arctanh}[\epsilon_{b}]$.
Thus
\begin{align}
\frac{\beta_{b} Q_{\text{tot}}(N)}{-\Delta S_{t}} = 4{\rm arctanh}[\epsilon_{b}]\frac{
\left(1-r^N\right)\left(\epsilon_\infty-\epsilon_{b}\right)}{\sum_{j=1}^{\infty}\frac{\epsilon_{N}^{2j}-\epsilon_{b}^{2j}}{j\left(2j-1\right)}}\geq 4/3,
\end{align}
which satisfies the Landauer principle but not giving the tight bound.

For the low polarization regime, this expression is simplified to 
\begin{align}
\lim_{\epsilon_b\to 0, N \to \infty}
    \frac{\beta_{b} Q_{\text{tot}}(N)}{-\Delta S_{t}} = \frac{4}{3}.
\end{align}


\section{Detailed evolution of HBACs using transition matrices}\label{sec_appen_calc}

\subsection{Calculation of PPA3-HBAC}

A round of PPA3 consists of applying the 3-qubit entropy compression $U_{\ppa}$ (given in \cref{eqn_U_PPA3_comp}), and thermalizing both machine qubits to a given bath temperature.
The thermalization can be effectively calculated as tracing out qubit 2 and 3, and replacing with two thermal qubits at the bath temperature.
The transition matrices representing these two operations are
\begin{equation}
    G_{tr23} = \begin{pmatrix}
        1 & 1 & 1 & 1 & 0 & 0 & 0 & 0 \\
        0 & 0 & 0 & 0 & 1 & 1 & 1 & 1
    \end{pmatrix} = \begin{pmatrix}
        \vec{1}_{1 \times 4} & \vec{1}_{0 \times 4}\\
        \vec{0}_{1 \times 4} & \vec{1}_{1 \times 4}
    \end{pmatrix},
\end{equation}

\begin{equation}
    G_{\otimes \rho_{b}^{\otimes 2}} = \begin{pmatrix}
        p_{b}^2 & 0 \\ p_{b} (1-p_{b}) & 0 \\ (1-p_{b}) p_{b} & 0 \\ (1-p_{b})^2 & 0 \\ 0 & p_{b}^2 \\ 0 & p_{b} (1-p_{b}) \\ 0 & (1-p_{b}) p_{b} \\ 0 & (1-p_{b})^2
    \end{pmatrix} \coloneqq \begin{pmatrix}
        \vec{s}_{b,2} & \vec{0}_{4 \times 1}\\
        \vec{0}_{4 \times 1} & \vec{s}_{b,2}
    \end{pmatrix},
\end{equation}
where we defined
\begin{equation}
    \vec{s}_{b,2} = \begin{pmatrix}
    p_{b}^2 \\ p_{b} (1-p_{b}) \\ (1-p_{b}) p_{b} \\ (1-p_{b})^2
    \end{pmatrix}
\end{equation}
to be the vector representing the state of two independent qubits both at the bath temperature ($p_{b}$ is the probability of being in $0$ for a thermal bath qubit at bath temperature).
The full transition matrix is given by
\begin{equation}
\begin{aligned}
    &G_{\ppa}^{\text{relax}} = G_{\otimes \rho_{b}^{\otimes 2}} G_{tr23}\\ 
    &= \begin{pmatrix}
    \vec{s}_{b,2} & \vec{s}_{b,2} & \vec{s}_{b,2} & \vec{s}_{b,2} & \vec{0}_{4 \times 1} & \vec{0}_{4 \times 1} & \vec{0}_{4 \times 1} & \vec{0}_{4 \times 1}\\
    \vec{0}_{4 \times 1} & \vec{0}_{4 \times 1} & \vec{0}_{4 \times 1} & \vec{0}_{4 \times 1} & \vec{s}_{b,2} & \vec{s}_{b,2} & \vec{s}_{b,2} & \vec{s}_{b,2} 
\end{pmatrix}.
\end{aligned}
\end{equation}

Combining with the transition matrix for the compression step,
\begin{equation}
    G_{\ppa}^{\text{comp}} = \begin{pmatrix}
        1 & 0 & 0 & 0 & 0 & 0 & 0 & 0\\
        0 & 1 & 0 & 0 & 0 & 0 & 0 & 0\\
        0 & 0 & 1 & 0 & 0 & 0 & 0 & 0\\
        0 & 0 & 0 & 0 & 1 & 0 & 0 & 0\\
        0 & 0 & 0 & 1 & 0 & 0 & 0 & 0\\
        0 & 0 & 0 & 0 & 0 & 1 & 0 & 0\\
        0 & 0 & 0 & 0 & 0 & 0 & 1 & 0\\
        0 & 0 & 0 & 0 & 0 & 0 & 0 & 1
    \end{pmatrix},
\end{equation}
the transition matrix for a complete round of PPA3 is
\begin{equation}
\begin{aligned}
    &G_{\ppa} = G_{\ppa}^{\text{relax}} G_{\ppa}^{\text{comp}}\\
    &= \begin{pmatrix}
        \vec{s}_{b,2} & \vec{s}_{b,2} & \vec{s}_{b,2}& \vec{0}_{4 \times 1}  & \vec{s}_{b,2} & \vec{0}_{4 \times 1} & \vec{0}_{4 \times 1} & \vec{0}_{4 \times 1}\\
        \vec{0}_{4 \times 1} & \vec{0}_{4 \times 1} & \vec{0}_{4 \times 1} & \vec{s}_{b,2} & \vec{0}_{4 \times 1} & \vec{s}_{b,2} & \vec{s}_{b,2} & \vec{s}_{b,2} 
    \end{pmatrix}.
\end{aligned}
\end{equation}
Thus, the post-cooling state after repeated rounds of PPA3 can be computed by simply raising $G_{\ppa}$ to the desired power, and multiplying it to the initial state vector.

Focusing on the target qubit alone, the evolution in a given round is first introducing two qubits at the bath temperature, apply the compression, then tracing out the two ancillary qubits.
The reduced transition matrix on the target qubit is given by
\begin{equation}
\begin{aligned}
    G_{\ppa, t} &= G_{tr23} G_{\ppa}^{\text{comp}} G_{\otimes \rho_{b}^{\otimes 2}}\\
    &= \begin{pmatrix}
    2 p_{b} - p_{b}^2 & p_{b}^2 \\ (1-p_{b})^2 & 1-p_{b}^2
    \end{pmatrix}.
\end{aligned}
\end{equation}
This provides a simpler view on how the target evolves during cooling.
Diagonalizing this matrix gives
\begin{equation}
    G_{\ppa, t} = \begin{pmatrix}
    z & -1 \\ 1 & 1
    \end{pmatrix} \begin{pmatrix}
    1 & 0 \\ 0 & y
    \end{pmatrix} \frac{1}{1+z} \begin{pmatrix}
    1 & 1 \\ -1 & z
    \end{pmatrix}
\end{equation}
where $y \coloneqq  2p_{b}(1-p_{b})$ and $z \coloneqq p_{b}^{2}/(1-p_{b})^{2}$.
We can raise it to the $N$-th power by
\begin{equation}
\begin{aligned}
    (G_{\ppa, t})^{N} &= \begin{pmatrix}
    z & -1 \\ 1 & 1
    \end{pmatrix} \begin{pmatrix}
    1 & 0 \\ 0 & y^{N}
    \end{pmatrix} \frac{1}{1+z} \begin{pmatrix}
    1 & 1 \\ -1 & z
    \end{pmatrix}
\end{aligned}
\end{equation}
Therefore, the state of the target after $N$ rounds starting from the initial state $\dket{\rho_{\ppa,t}(0)} = (p_{t},1-p_{t})^{T}$ is
\begin{equation}\label{eqn_appen_ppa3_reduced_state}
\begin{aligned}
    &\dket{\rho_{\ppa,t}(N)} = (G_{\ppa, t})^{N} \dket{\rho_{\ppa,t}(0)}\\
    &= \frac{1}{1+z} \begin{pmatrix}
        z-y^{N}(z-(z+1)p_{t})\\
        1+y^{N}(z-(z+1)p_{t})
    \end{pmatrix}.
\end{aligned}
\end{equation}
for $N \geq 1$. Since $y \in [0,1/2]$, it is easy to obtain the asymptotic state as
\begin{equation}
\begin{aligned}
    \dket{\rho_{\ppa,t}(\infty)} &= \lim_{N \rightarrow \infty} \dket{\rho_{\ppa,t}(N)} = \frac{1}{1+z} \begin{pmatrix}
    z \\ 1
    \end{pmatrix}.
\end{aligned}
\end{equation}

We now switch back to using the full transition matrix to compute thermodynamic quantities for calculating cooling efficiencies.
The work input for the $N$-th round is the energy increase for the full system during the compression step,
\begin{equation}\label{eqn_appen_W_ppa3}
\begin{aligned}
    &w_{\ppa}(N) = \dbra{H_{\ppa}} (G_{\ppa}^{\text{comp}} - \mathds{1}) \dket{\rho_{\ppa}(N-1)}\\
    &= \dbra{H_{\ppa}} (G_{\ppa}^{\text{comp}} - \mathds{1}) (G_{\ppa})^{N-1} \dket{\rho_{\ppa}(0)}.
\end{aligned}
\end{equation}
The vectorized Hamiltonian is
\begin{equation}
    \dbra{H_{\ppa}} = \sum_{i=1}^{3} \dbra{\omega_{i} \ketbra{1}{1}_{[i]}} = \omega \begin{pmatrix}
        0 & 1 & 1 & 2 & 1 & 2 & 2 & 3 
    \end{pmatrix}
\end{equation}
where we assumed the same energy gap $\omega$ for each qubit.
The vectorized initial state is
\begin{equation}
    \dket{\rho_{\ppa}(0)} = \dket{\bigotimes_{i=1}^{3} \rho_{i}} = \begin{pmatrix}
        p_{b}^{3} \\ p_{b}^{2}(1-p_{b}) \\ p_{b}^{2}(1-p_{b}) \\ p_{b}(1-p_{b})^{2} \\ p_{b}^{2}(1-p_{b}) \\ p_{b}(1-p_{b})^{2} \\ p_{b}(1-p_{b})^{2} \\ (1-p_{b})^{3} 
    \end{pmatrix}
\end{equation}
where we assumed that all qubits start from the same (bath) temperature for simplicity.
Directly calculating \cref{eqn_appen_W_ppa3} using a computational software gives
\begin{equation}
    w_{\ppa}(N) = \omega (p_{b}-\frac{1}{2}) y^{N},\  N \geq 1
\end{equation}

Meanwhile, the energy decrease for the target during the $N$-th round can be computed from \cref{eqn_appen_ppa3_reduced_state}: 
\begin{equation}
\begin{aligned}
    -\Delta e_{t}(N) &= \dbra{H_{1}} (\dket{\rho_{\ppa,t}(N-1)} - \dket{\rho_{\ppa,t}(N)})\\
    &= \omega (p_{b}-\frac{1}{2}) y^{N},\ N \geq 1
\end{aligned}
\end{equation}
It can be seen from here that both the per-round and cumulative CoP are simply
\begin{equation}
    k_{\ppa}(N) =  K_{\ppa}(N) = 1.
\end{equation}

\subsection{Calculation of NOE2-HBAC}
A round of NOE-HBAC consists of first applying the CMS operation on the second qubit, then applying the $\Gamma_{2}$ relaxation.
The transition matrices representing these two operations are
\begin{equation}
    G_{\text{CMS}} = \frac{1}{2} \begin{pmatrix}
        1 & 1 & 0 & 0 \\
        1 & 1 & 0 & 0 \\
        0 & 0 & 1 & 1 \\
        0 & 0 & 1 & 1
    \end{pmatrix},
\end{equation}

\begin{equation}
    G_{\Gamma_{2}} = \begin{pmatrix}
        p_{b2} & 0 & 0 & p_{b2} \\
        0 & 1 & 0 & 0 \\
        0 & 0 & 1 & 0 \\
        1-p_{b2} & 0 & 0 & 1-p_{b2}
    \end{pmatrix},
\end{equation}
where $p_{b2}$ is defined in \cref{eqn_defn_p2}.
A round of NOE-HBAC is represented by the transition matrix
\begin{equation}
\begin{aligned}
    &G_{\noe} = G_{\Gamma_{2}} G_{\text{CMS}}\\
    &= \frac{1}{2} \begin{pmatrix}
        p_{b2} & p_{b2} & p_{b2} & p_{b2} \\
        1 & 1 & 0 & 0 \\
        0 & 0 & 1 & 1 \\
        1-p_{b2} & 1-p_{b2} & 1-p_{b2} & 1-p_{b2}
    \end{pmatrix}.
\end{aligned}
\end{equation}

The transition matrix representing the evolution of the target qubit is
\begin{equation}
\begin{aligned}
    G_{\noe, t} &= G_{tr2} G_{\Gamma_{2}} G_{\otimes \rho_{\text{CMS}}}\\
    &= \frac{1}{2} \begin{pmatrix}
    p_{b2} +1 & p_{b2} \\ 1-p_{b2} & 2-p_{b2}
    \end{pmatrix}.
\end{aligned}
\end{equation}
Let $x \coloneqq \frac{p_{b2}}{1-p_{b2}}$; then $G_{\noe, t}$ can be diagonalized as
\begin{equation}
    G_{\noe, t} = \begin{pmatrix}
    x & -1 \\ 1 & 1
    \end{pmatrix}  \begin{pmatrix}
    1 & 0 \\ 0 & \frac{1}{2}
    \end{pmatrix} \frac{1}{1+x} \begin{pmatrix}
    1 & 1 \\ -1 & x
    \end{pmatrix}
\end{equation}
Raising to the $N$-th power:
\begin{equation}
    G_{\noe, t}^{N} = \frac{1}{1+x} \begin{pmatrix}
    x + (\frac{1}{2})^{N} & x(1-(\frac{1}{2})^{N}) \\ 1-(\frac{1}{2})^{N} & 1+(\frac{1}{2})^{N} x
    \end{pmatrix}.
\end{equation}
The state of the target after $N$ rounds starting from the initial state $\dket{\rho_{t(\noe)}(0)} = (p_{t},1-p_{t})^{T}$ is
\begin{equation}\label{eqn_appen_noe_reduced_state}
\begin{aligned}
    &\dket{\rho_{t(\noe)}(N)} = (G_{\noe, t})^{N} \dket{\rho_{t(\noe)}(0)}\\
    &= \frac{1}{1+x} \begin{pmatrix}
        x+2^{-N}(p_{t}-x+p_{t} x)\\
        1-2^{-N}(p_{t}-x+p_{t} x)
    \end{pmatrix}
\end{aligned}
\end{equation}
for $N \geq 1$.
The asymptotic state is
\begin{equation}
    \dket{\rho_{t(\noe)}(\infty)} = \lim_{N \rightarrow \infty} \dket{\rho_{t(\noe)}(N)} = \frac{1}{1+x} \begin{pmatrix}
    x \\ 1
    \end{pmatrix}.
\end{equation}

The work input for the $N$-th round is the energy increase for the full system during the CMS step,
\begin{equation}\label{eqn_appen_W_noe}
\begin{aligned}
    &w_{\noe}(N) = \dbra{H_{\noe}} (G_{\text{CMS}} - \mathds{1}) \dket{\rho_{\noe}(N-1)}\\
    &= \dbra{H_{\noe}} (G_{\text{CMS}} - \mathds{1}) (G_{\noe})^{N-1} \dket{\rho_{\noe}(0)}.
\end{aligned}
\end{equation}
The vectorized Hamiltonian is
\begin{equation}
    \dbra{H_{\noe}} = \sum_{i=1}^{2} \dbra{\omega_{i} \ketbra{1}{1}_{[i]}} = \omega \begin{pmatrix}
        0 & 1 & 1 & 2
    \end{pmatrix}
\end{equation}
where we assumed the same energy gap $\omega$ for each qubit.
The vectorized initial state is
\begin{equation}
    \dket{\rho_{\noe}(0)} = \dket{\bigotimes_{i=1}^{2} \rho_{i}} = \begin{pmatrix}
        p_{b}^{2} \\ p_{b}(1-p_{b}) \\ p_{b}(1-p_{b}) \\ (1-p_{b})^{2}
    \end{pmatrix}
\end{equation}
where we assumed that all qubits start from the same (bath) temperature for simplicity.
Directly calculating \cref{eqn_appen_W_noe} using a computational software gives
\begin{equation}
    w_{\noe}(N) = \begin{cases}
        \omega (p_{b} - \frac{1}{2}), & N=1\\
         \omega (p_{b2} - p_{b}) 2^{1-N}, & N>1 
    \end{cases}
\end{equation}
The cumulative work up to round $N$ is
\begin{equation}
    W_{\noe}(N) = \sum_{i=1}^{N} w_{\noe}(i) = \omega ((p_{b2} - \frac{1}{2}) - 2^{1-N} (p_{b2} - p_{b})).
\end{equation}

Meanwhile, the energy decrease for the target during the $N$-th round can be computed from \cref{eqn_appen_noe_reduced_state}: 
\begin{equation}
\begin{aligned}
    -\Delta e_{t}(N) &= \dbra{H_{1}} (\dket{\rho_{t(\noe)}(N-1)} - \dket{\rho_{t(\noe)}(N)})\\
    &= \omega (p_{b2} - p_{b}) 2^{-N}
\end{aligned}
\end{equation}
for all $N \geq 1$.
The total energy decrease up to round $N$ is
\begin{equation}
    -\Delta E_{t}(N) = \sum_{i=1}^{N} -\Delta e_{t}(i) = (p_{b2}-p_{b}) (1-2^{-N}) \omega.
\end{equation}

The COP for the $N$-th round is
\begin{equation}
    k_{\noe}(N) = \frac{-\Delta e_{t(\noe)}(N)}{w_{\noe}(N)} = \begin{cases}
        \frac{p_{b2}-p_{b}}{2 p_{b}-1}, & N=1 \\
        \frac{1}{2}, & N>1 
    \end{cases}
\end{equation}
and the cumulative CoP up to round $N$ is
\begin{equation}
    K_{\noe}(N) = \frac{-\Delta E_{t(\noe)}(N)}{W_{\noe}(N)} = \frac{(p_{b2}-p_{b}) (1-2^{-N})}{(p_{b2} - \frac{1}{2}) - 2^{1-N} (p_{b2} - p_{b})}.
\end{equation}

\subsection{Calculation of SR2-HBAC}
A round of SR2-HBAC consists of first applying the  $\sigma_{x}$ gate on the second qubit, then applying the $\Gamma_{2}$ relaxation, and finally thermalizing the second qubit.
The transition matrices are
\begin{equation}
    G_{\sigma_{x},[2]} = \begin{pmatrix}
        0 & 1 & 0 & 0 \\
        1 & 0 & 0 & 0 \\
        0 & 0 & 0 & 1 \\
        0 & 0 & 1 & 0
    \end{pmatrix},
\end{equation}
\begin{equation}
    G_{\Gamma_{1},[2]} = \begin{pmatrix}
        p_{b} & p_{b} & 0 & 0 \\
        1-p_{b} & 1-p_{b} & 0 & 0 \\
        0 & 0 & p_{b} & p_{b} \\
        0 & 0 & 1-p_{b} & 1-p_{b}
    \end{pmatrix},
\end{equation}
and $G_{\Gamma_{2}}$ as defined as above.
A round of SR2-HBAC is represented by the transition matrix
\begin{equation}
\begin{aligned}
    &G_{\srg} = G_{\Gamma_{1},[2]} G_{\Gamma_{2}} G_{\sigma_{x},[2]}=\\
    &\begin{pmatrix}
        p_{b} & p_{b} p_{b2} & p_{b} p_{b2} & 0 \\
        1-p_{b} & (1-p_{b}) p_{b2} & (1-p_{b}) p_{b2} & 0 \\
        0 & p_{b}(1-p_{b2}) & p_{b}(1-p_{b2}) & p_{b} \\
        0 & (1-p_{b})(1-p_{b2}) & (1-p_{b})(1-p_{b2}) & 1-p_{b}
    \end{pmatrix}.
\end{aligned}
\end{equation}

The transition matrix representing the evolution of the target qubit is
\begin{equation}
\begin{aligned}
    G_{\srg, t} &= G_{tr2} G_{\Gamma_{2}} G_{\sigma_{x},[2]} G_{\otimes \rho_{b}}\\
    &= \begin{pmatrix}
    p_{b} +(1-p_{b})p_{b2} & p_{b} p_{b2} \\ (1-p_{b})(1-p_{b2}) & 1-p_{b} p_{b2}
    \end{pmatrix}.
\end{aligned}
\end{equation}
Let $w \coloneqq p_{b2} p_{b}$, $v = 1-p_{b2} -p_{b} + p_{b2} p_{b}$, $u = 1-v-w$, then $G_{\srg, t}$ can be diagonalized as
\begin{equation}
    G_{\srg, t} = \begin{pmatrix}
    \frac{w}{v} & -1 \\ 1 & 1
    \end{pmatrix} \begin{pmatrix}
    1 & 0 \\ 0 & u
    \end{pmatrix} \frac{v}{w+v} \begin{pmatrix}
    1 & 1 \\ -1 & \frac{w}{v}
    \end{pmatrix}
\end{equation}
Raising to the power of $N$:
\begin{equation}
    G_{\srg, t}^{N} = \frac{1}{w+v} \begin{pmatrix}
    w + v u^{N} & w - w u^{N} \\ v - v u^{N} & v + w u^{N}
    \end{pmatrix}.
\end{equation}
The state of the target after $N$ rounds starting from the initial state $\dket{\rho_{t(\srg)}(0)} = (p_{t},1-p_{t})^{T}$ is
\begin{equation}\label{eqn_appen_srg_reduced_state}
\begin{aligned}
    &\dket{\rho_{t(\srg)}(N)} = (G_{\srg, t})^{N} \dket{\rho_{t(\srg)}(0)}\\
    &= \frac{1}{w+v} \begin{pmatrix}
        w+u^{N}(-w+p_{t}(v+w))\\
        v-u^{N}(-w+p_{t}(v+w))
    \end{pmatrix}
\end{aligned}
\end{equation}
for $N \geq 1$.
The asymptotic state is
\begin{equation}
\begin{aligned}
    \dket{\rho_{t(\srg)}(\infty)} &= \lim_{N \rightarrow \infty} \dket{\rho_{t(\srg)}(N)} = \frac{1}{w+v} \begin{pmatrix}
    w \\ v
    \end{pmatrix}
\end{aligned}
\end{equation}
since $\abs{u} < 1$.

The work input for the $N$-th round is the energy increase for the full system during the $\sigma_{x}$ step,
\begin{equation}\label{eqn_appen_W_srg}
\begin{aligned}
    &W_{\srg}(N) = \dbra{H_{\srg}} (G_{\sigma_{x},[2]} - \mathds{1}) \dket{\rho_{\srg}(N-1)}\\
    &= \dbra{H_{\srg}} (G_{\sigma_{x},[2]} - \mathds{1}) (G_{\srg})^{N-1} \dket{\rho_{\srg}(0)}.
\end{aligned}
\end{equation}
The vectorized Hamiltonian is again
\begin{equation}
    \dbra{H_{\srg}} = \sum_{i=1}^{2} \dbra{\omega_{i} \ketbra{1}{1}_{[i]}} = \omega \begin{pmatrix}
        0 & 1 & 1 & 2
    \end{pmatrix}
\end{equation}
where we assumed the same energy gap $\omega$ for each qubit.
The vectorized initial state is
\begin{equation}
    \dket{\rho_{\srg}(0)} = \dket{\bigotimes_{i=1}^{2} \rho_{i}} = \begin{pmatrix}
        p_{b}^{2} \\ p_{b}(1-p_{b}) \\ p_{b}(1-p_{b}) \\ (1-p_{b})^{2}
    \end{pmatrix}
\end{equation}
where we assumed that all qubits start from the same (bath) temperature for simplicity.
Directly calculating \cref{eqn_appen_W_srg} using a computational software gives
\begin{equation}
    w_{\srg}(N) = \omega (2p_{b}-1)
\end{equation}
for all $N \geq 1$.
In fact, this can be trivially obtained by noticing that the $\sigma_{x}$ gate always flips the same thermal state at bath temperature in each round.
The cumulative work up to round $N$ is
\begin{equation}
    W_{\srg}(N) = N \omega (2p_{b}-1)
\end{equation}

Meanwhile, the energy decrease for the target during the $N$-th round can be computed from \cref{eqn_appen_srg_reduced_state}: 
\begin{equation}
\begin{aligned}
    -\Delta e_{t(\srg)}(N) &= \dbra{H_{1}} (\dket{\rho_{t(\srg)}(N-1)} - \dket{\rho_{t(\srg)}(N)})\\
    &= \omega p_{b} (1 - p_{b}) (2p_{b2}-1) u^{N-1}
\end{aligned}
\end{equation}
for all $N \geq 1$.
The total energy decrease in the target is 
\begin{equation}
    -\Delta E_{t(\srg)}(N) = \omega p_{b} (1 - p_{b}) (2p_{b2}-1) \frac{1-u^{N}}{1-u}
\end{equation}
The COP for the $N$-th round is
\begin{equation}
    k_{\srg}(N) = \frac{-\Delta e_{t(\srg)}(N)}{w_{\srg}(N)} = \frac{p_{b} (1 - p_{b}) (2p_{b2}-1)}{2p_{b}-1} u^{N-1}
\end{equation}
for all rounds $N \geq 1$.
The cumulative CoP up to round $N$ is
\begin{equation}
\begin{aligned}
    K_{\srg}(N) &= \frac{-\Delta E_{t(\srg)}(N)}{W_{\srg}(N)} \\
    &= \frac{p_{b} (1 - p_{b}) (2p_{b2}-1)}{2p_{b}-1} \frac{1-u^{N}}{N(1-u)}.
\end{aligned}
\end{equation}

\subsection{Calculation of xHBAC1}
A round of 1-qubit xHBAC consists of first applying the $\sigma_{x}$ gate on the target, then applying the $\beta$-swap operation.
The transition matrix for $\beta$-swap is
\begin{equation}
    G_{\beta} = \begin{pmatrix}
    1-\omega e^{-\beta_{b}}  & 1 \\ e^{-\beta_{b} \omega} & 0
    \end{pmatrix}
\end{equation}
where $\omega$ is the energy gap of the qubit and $\beta_{b}$ is the bath inverse temperature.
The total $G$-matrix is given by
\begin{equation}
    G_{\xhbac} = G_{\beta} G_{\sigma_{x}} = \begin{pmatrix}
    1 & 1-\omega e^{-\beta_{b}} \\ 0 & e^{-\beta_{b} \omega}
    \end{pmatrix}.
\end{equation}
To diagonalize $G_{\xhbac}$, let $s=e^{-\beta_{b} \omega}$, we obtain
\begin{equation}
    G_{\xhbac} = \begin{pmatrix}
    1 & -1 \\ 0 & 1
    \end{pmatrix} \begin{pmatrix}
    1 & 0 \\ 0 & s
    \end{pmatrix} \begin{pmatrix}
    1 & 1 \\ 0 & 1
    \end{pmatrix}.
\end{equation}
The $N$-th power of $G_{\xhbac}$ is given by
\begin{equation}
    G_{\xhbac}^{N} = \begin{pmatrix}
    1 & 1-s^{N} \\ 0 & s^{N}
    \end{pmatrix}
\end{equation}
The state of the target after $N$ rounds starting from the initial state $\dket{\rho_{t(\xhbac)}(0)} = (p_{t},1-p_{t})^{T}$ is
\begin{equation}\label{eqn_appen_xhbac_reduced_state}
\begin{aligned}
    \dket{\rho_{t(\xhbac)}(N)} &= (G_{\xhbac, t})^{N} \dket{\rho_{t(\xhbac)}(0)}\\
    &= \begin{pmatrix}
        1- s^{N}(1-p_{b})\\
        s^{N} (1-p_{b})
    \end{pmatrix}
\end{aligned}
\end{equation}
for $N \geq 1$.
The asymptotic state is
\begin{equation}
    \dket{\rho_{t(\xhbac)}(\infty)} = \lim_{N \rightarrow \infty} \dket{\rho_{t(\xhbac)}(N)} = \begin{pmatrix}
    1 \\ 0
    \end{pmatrix},
\end{equation}
i.e., the pure state $\ket{0}$.

The work input for the $N$-th round is the energy increase for the full system during the $\sigma_{x}$ step,
\begin{equation}\label{eqn_appen_W_xhbac}
\begin{aligned}
    &w_{\xhbac}(N) = \dbra{H_{\xhbac}} (G_{\sigma_{x}} - \mathds{1}) \dket{\rho_{\xhbac}(N-1)}\\
    &= \dbra{H_{\xhbac}} (G_{\sigma_{x}} - \mathds{1}) (G_{\xhbac})^{N-1} \dket{\rho_{\xhbac}(0)}\\
    &= \omega(1-2s^{N-1}(1-p_{b})).
\end{aligned}
\end{equation}
The total work input is
\begin{equation}
    W_{\xhbac}(N) = \omega \left(n-2(1-p_{b}) \frac{1-s^{N}}{1-s} \right).
\end{equation}
The energy decrease for the target during the $N$-th round is
\begin{equation}
\begin{aligned}
    -\Delta e_{t(\xhbac)}(N) &\\
   = \dbra{H_{1}} &(\dket{\rho_{\xhbac}(N-1)} - \dket{\rho_{\xhbac}(N)})\\
    = \omega (1 - & p_{b}) (1-s) s^{N-1}
\end{aligned}
\end{equation}
for all $N \geq 1$.
The total energy decrease up to round $N$ is
\begin{equation}
    -\Delta E_{t(\xhbac)}(N) = \omega(1-p_{b})(1-s^{N}).
\end{equation}
The COP for the $N$-th round is
\begin{equation}
    k_{\xhbac}(N) = \frac{-\Delta e_{t}(N)}{w_{\xhbac}(N)} = \frac{(1 - p_{b})(1-s)}{1-2s^{N-1}(1-p_{b})} s^{N-1}
\end{equation}
for all rounds $N \geq 1$.
The cumulative COP up to round $N$ is
\begin{equation}
    K_{\xhbac}(N) = \frac{(1-p_{b})(1-s)(1-s^{N})}{n(1-s)-2(1-p_{b}) (1-s^{N})}
\end{equation}

\section{Proof of the equilibrium state of the improved PPA}\label{appen_proof_efficient_PPA}
In the improved PPA, the goal is to minimize the work input during the compression step, while keeping the same final temperature of the target qubit as in PPA.
This is achieved by placing the diagonal elements in the $\ket{0}_{t}$ subspace of the target qubit in the lowest energy levels in decreasing order, and similarly for the $\ket{1}_{t}$ subspace.

In every cycle, the system will start from an initial tensor product state where the first qubit is in the state with diagonal elements $(1-\delta_{t}, \delta_{t})$, and all other qubits have diagonal elements $(1-\delta_{b}, \delta_{b})$, where $\delta_{b} = \bra{1} \rho_{b}^{th} \ket{1}$.
The full state for $n$ qubits has $2^{n}$ diagonal elements, which can be grouped by the Hamming weight of their corresponding state.
For example, the population corresponding to the state $\ket{00101}$, which has weight $2$, is given by $(1-\delta_{t})(1-\delta_{b})^{2} \delta_{b}^{2}$.
In general, for a total weight of $i$, the diagonal elements are
\begin{equation}
    \begin{gathered}
    \binom{n-1}{i}:\ \delta_{0}(i) = (1-\delta_{t}) (1-\delta_{b})^{n-i-1} \delta_{b}^{i},\ i=0 \dots (n-1)\\
    \binom{n-1}{i-1}:\ \delta_{1}(i) = \delta_{t} (1-\delta_{b})^{n-i} \delta_{b}^{i-1},\ i=1 \dots n
    \end{gathered}
\end{equation}
where the binomial coefficients denote how many identical terms there are in total, $\delta_{0}(i)$ corresponds to what's originally in the $\ket{0}_{t}$ subspace and the $\delta_{1}(i)$ in the $\ket{1}_{t}$ subspace of the target qubit.

Consider the situation where the target starts from a state with a $\delta_{t}$ satisfying the following relation:
\begin{equation}\label{eqn_appen_equil_EPPA}
    \frac{1-\delta_{t}^{*}}{\delta_{t}^{*}} = \left(\frac{1-\delta_{b}}{\delta_{b}} \right)^{n-1}.
\end{equation}
We can now rewrite $\delta_{0}(i)$ and $\delta_{1}(i)$ as
\begin{equation}
    \begin{aligned}
        &\left\{\delta_{0}(i), \delta_{1}(i) \right\}\\
        &= \left\{\left(\frac{1-\delta_{b}}{\delta_{b}} \right)^{n-1} (1-\delta_{b})^{n-i-1} \delta_{b}^{i} \delta_{t}^{*}, \delta_{t}^{*} (1-\delta_{b})^{n-i} \delta_{b}^{i-1} \right\}\\
        &= \delta_{t}^{*} (1-\delta_{b})^{n-1} \left\{ \left(\frac{1-\delta_{b}}{\delta_{b}} \right)^{n-i-1},\ \left(\frac{1-\delta_{b}}{\delta_{b}} \right)^{-i+1} \right\}
,    \end{aligned}
\end{equation}
where we have factored out the common term $\delta_{t}^{*} (1-\delta_{b})^{n-1}$.
The fraction $(1-\delta_{b})/\delta_{b}$ is the ratio of $\ket{0}$ over $\ket{1}$ populations for a qubit at the bath temperature and is always $\geq 1$ for all $\delta_{b} \in (0, 1/2]$.
Therefore, each term increases with an increasing exponent.
Specifically, since both exponents decrease with $i$, the \textit{smallest} term from the $\ket{0}_{t}$ subspace is achieved at $i=n-1$, and the \textit{largest} term from the $\ket{1}_{t}$ subspace is achieved at $i=1$.
Since
\begin{equation}
    n-(n-1)-1 = -(1)-1 = 0,
\end{equation}
these two terms are in fact equal to each other.
This implies that all the largest $2^{n-1}$ terms are already in the $\ket{0}_{t}$ subspace, so the temperature of the target qubit will remain the same after the compression step.
And since the target's initial temperature is the only variable during the compression, we conclude that this state is a stable fixed point for the E-PPA protocol.

Finally, we see that the equilibrium population satisfying \cref{eqn_appen_equil_EPPA} has an inverse temperature
\begin{equation}
\begin{aligned}
    \beta_{b} \omega &= \log(\left(\frac{1-\delta_{b}}{\delta_{b}} \right)^{n-1}) = (n-1) \log\left(\frac{1-\delta_{b}}{\delta_{b}} \right)\\ 
    &= (n-1) \beta_{b} \omega
\end{aligned}
\end{equation}
where $\beta_{b}$ is the bath's inverse temperature.
Therefore, the equilibrium temperature is $T_{b}/(n-1)$.

\bibliographystyle{apsrev4-2}
\bibliography{AC}

\end{document}